\documentclass[sigconf]{acmart}

\AtBeginDocument{%
  \providecommand\BibTeX{{%
    \normalfont B\kern-0.5em{\scshape i\kern-0.25em b}\kern-0.8em\TeX}}}

\copyrightyear{2023}
\acmYear{2023}
\setcopyright{rightsretained}
\acmConference[DIS '23]{Designing Interactive Systems Conference}{July 10--14, 2023}{Pittsburgh, PA, USA}
\acmBooktitle{Designing Interactive Systems Conference (DIS '23), July 10--14, 2023, Pittsburgh, PA, USA}
\acmDOI{10.1145/3563657.3595982}
\acmISBN{978-1-4503-9893-0/23/07}

\usepackage{graphicx}
\usepackage{float}
\usepackage{multirow}
\usepackage{soul}

\begin{document}

\title{Co-Designing Alternatives for the Future of Gig Worker Well-Being
\subtitle{Navigating Multi-Stakeholder Incentives and Preferences}}

\author{Jane Hsieh}
\email{jhsieh2@cs.cmu.edu}
\affiliation{%
  \institution{Carnegie Mellon University}
  \streetaddress{5000 Forbes Ave}
  \city{Pittsburgh}
  \state{PA}
  \country{USA}
  \postcode{15213}
}

\author{Miranda Karger}
\email{mak2307@barnard.edu}
\affiliation{%
  \institution{Barnard College}
  \city{New York City}
  \state{New York}
  \country{USA}
}

\author{Lucas Zagal}
\email{u1051220@utah.edu}
\affiliation{%
  \institution{University of Utah}
  \city{Salt Lake City}
  \state{Utah}
  \country{USA}
}
\author{Haiyi Zhu}
\email{haiyiz@cs.cmu.edu}
\affiliation{%
  \institution{Carnegie Mellon University}
  \streetaddress{5000 Forbes Ave}
  \city{Pittsburgh}
  \state{PA}
  \country{USA}
  \postcode{15213}
}

\renewcommand{\shortauthors}{Hsieh et al.}

\newcommand{\add}[1]{\textcolor{red}{#1}} 
\newcommand{\delete}[1]{{\textcolor{red}{\sout{#1}}}}
\renewcommand{\delete}[1]{}
\renewcommand{\add}[1]{#1}

\begin{abstract}
Gig workers, and the products and services they provide, play an increasingly ubiquitous role in our daily lives. But despite growing evidence suggesting that worker well-being in gig economy platforms have become significant societal problems, few studies have investigated possible solutions. We take a stride in this direction by engaging workers, platform employees, and local regulators in a series of speed dating workshops using storyboards based on real-life situations to rapidly elicit stakeholder preferences for addressing financial, physical, and social issues related to worker well-being. Our results reveal that existing public and platformic infrastructures fall short in providing workers with resources needed to perform gigs, surfacing a need for multi-platform collaborations, technological innovations, as well as changes in regulations, labor laws, and the public's perception of gig workers, among others. Drawing from multi-stakeholder findings, we discuss these implications for technology, policy, and service as well as avenues for collaboration.

\end{abstract}





\maketitle

\section{Introduction}
The rapid growth of the gig economy has motivated individuals around the globe to engage in more flexible and autonomous forms of work. Gig and platform-based work are presently characterized by short-term, on-demand work completed by independent contractors who get paid in return for the ``gigs'' they perform. Upon first glance, digital labor platforms seem to benefit everyone involved, offering workers novel job opportunities, enabling small businesses to scale quickly, and providing individual consumers services like ridesharing and food delivery \cite{Blaising2021-ue,Wood2018-lq}. But under the surface, the amalgamation of low compensation, high competition, and the just-in-time nature of gig work leaves individual contractors toiling at odd hours for prolonged periods of time, and with insufficient compensation for making a living. Unlike employees of traditional firms, gig workers are not entitled to employee benefits such as healthcare or retirement contributions \cite{Bajwa2018-gy,Yao2021-eb}. With the proliferation of online gig platforms that facilitate short-term work, individual contractors increasingly experience competition, lowered wages, a commodification of labor, job precarity and in general adverse working conditions \cite{Blaising2021-ue,Dubal2017-bj,Berger2019-yt,Kalleberg_undated-sb,Wood2019-jp}. 

In appearance, the gig economy model offers workers flexibility and low-entry barriers while affording consumers time and cost savings. However, such conveniences are possible only because it ``necessitates cutting every cost possible, usually by externalizing them through misclassifying workers so they do not qualify for expensive benefits like a minimum wage or health insurance'' \cite{Luckett_undated-uq}. Prior works in the domain have extensively studied how such reduced working conditions negatively impact the well-being of gig and contractual workers \cite{schulze2022algorithmic, li2022well, arnoldi2021mapping}, induced by a lack of social, financial, technological and regulatory support \cite{Kalleberg_undated-sb,Almoqbel2019-in,Yao2021-eb,Wood2019-jp, glavin2021alienated}. For instance, in their seminal work examining job quality in gig work, Wood et. al. described how platformic control causes workers to have weak structural power compared to clients, which results in burnout \cite{Wood2018-lq}. Yao et. al. found that while social media groups enabled workers to share experiential knowledge amongst one another, they fell short in building a collective identity among workers since strategic information-sharing could harm an individual worker's comparative advantage \cite{Yao2021-eb}. Howard investigated how labor laws apply in non-standard gig work arrangements, underscoring the health and safety risks involved for workers in such environments \cite{Howard2017-wd}. 

Recent bodies of work within HCI increasingly urge and pursue the design of systems from a worker-centered perspective \cite{Yang2020-dg,zhang2022algmanagement,Park2022-qy}. As a first step in this direction, Zhang et. al. codesigned alternative platform futures with workers to minimize the impact of algorithmic management on well-being \cite{zhang2022algmanagement}. In their research agenda, Ashford et. al. drew from organizational behavior theory to delineate potential behaviors that individual workers can capitalize on to thrive in the new world of work \cite{Ashford2018-dw}. While these studies focus on worker-driven solutions, improving gig worker conditions requires the active involvement of and collaboration with multiple stakeholder groups \cite{Forlizzi2018-eb, goods2019your}. Expertise of regulators and lawmakers are required to craft and enforce mandates and labor regulations that govern the gig economy \cite{bates2021lessons, Dubal2019-qi}, support from platforms is crucial to implement programs and engage in co-regulation \cite{cannon2014framework, healy2020sceptics}, and worker input is indispensable to designing legal and platformic changes that engender practical and productive impact \cite{Johnston_undated-wy, lord2022sustainability}. 

Our work involved a diverse set of stakeholders, and by leveraging the speed dating method, we collaborated with participants from within the United States to brainstorm, develop and assess a wide range of service, policy and technological interventions for addressing the various social, financial and physical challenges of gig work \cite{Davidoff2007-fq}. The hidden costs and challenges emerging from such past bodies of work, combined with themes uncovered from local workshops and news articles, informed the construction of scenarios for our workshops. During the codesign sessions, speed dating allowed us to incorporate reported issues into scenarios accompanied by provocative questions and solutions, empowering us to 1.) learn latent social needs and boundaries of stakeholders and 2.) imagine and evaluate solutions without the high efforts of implementation. 
In conducting these workshops, we sought to answer the following research questions:

\textbf{Research questions: }
\begin{enumerate}
\item What incentives, preferences and deterrents do stakeholders have in supporting and implementing solutions for improving gig worker well-being?
\item What are the most desirable and feasible changes for improving challenges present in gig work?
\end{enumerate}

Our multi-stakeholder workshops allowed us to share key quantitative and qualitative insights from regulators, platform practitioners and the gig workforce at large, revealing details about shared worker struggles, desired benefits and steps that stakeholders can take to turn imagined futures into reality. Thus, we make unique research contributions by 1.) presenting improvements to the gig work condition that are acceptable to multiple stakeholders groups and 2.) offering a discussion of how stakeholders can contribute to solutions and interventions. Through this endeavor, we hope to contribute to a future gig workplace that tracks and improves workers' physical, financial and social well being, so as to approach more equitable and inclusive gig platforms and communities. 

\section{Related Works}
Gig work, at large, can be characterized as ``electronically mediated employment arrangements in which individuals find short-term tasks or projects via websites or mobile apps that connect them to clients and process payment'' \cite{Kuhn2019-fi}. However, further segmentation can divide gig work into app work (e.g. Uber, DoorDash, TaskRabbit), crowdwork (e.g. Amazon Mechanical Turk), and capital platform work (e.g. Airbnb, Etsy) \cite{Duggan2020-qh}. A similar categorization sections gig work into local and remote parts, with the former consisting of manual labor (e.g. transport, food delivery, furniture assembly) and the latter comprising of digital services such as software development or logo design \cite{Huws2016-vv}. At the start, we focused primarily on app workers performing physical tasks, but found capital platform workers to share many of the same risks and challenges after reviewing relevant literature and articles. Thus, our workshops aim to address the various social, financial and power struggles as well as health and physical risks endemic to these two forms of gig work. In the following, we summarize five major shortcomings of gig work explored in past studies that informed our workshop design.

\subsection{Risks and Challenges of Gig Work} \label{challenges}
\subsubsection{Missing Employment Benefits} 
Although gig work offers more flexible work hours, limited employment benefits forces workers to complete additional hours of unpaid labor \cite{anwar2022faux}.
While many workers prefer to keep their legal classification as independent contractors for the associated flexibilities (e.g. no employer attachments), the lack of a formal employment arrangement costs them many benefits and protections, including wage guarantees, workers' compensation, unemployment insurance, healthy and safe work spaces, and the right to unionization \cite{Dubal2019-qi}. The deprivation of workers' rights and protections that contractors experience (which especially impoverishes the mental health of working mothers \cite{kirwin2022working}) has been longstanding, with accounts dating back to at least 2002 \cite{ilo2002resolution}.

In an effort to avoid employment regulations, many gig platforms leverage workers' desires to remain contractors as an argument in court to avoid responsibilities of providing employee benefits. This argument for platforms is frequently used in trials since as early as 2017, after which more than 100 such US lawsuits have been filed against Uber regarding driver misclassifications, with many more appearing across other platforms and nations \cite{de2021platform, Bales_undated-ok}. To continue exploiting the legal loophole in employment classifications, gig platforms have spent hundreds of millions to lobby for the ballot measure Prop 22 in the summer of 2021 \cite{conger2021judge}. Presently, how workers should be classified remains an ongoing debate -- the control and economic realities tests that serve to distinguish between employees and independent contractors both lead to indeterminate results when applied to rideshare drivers, and different courts'  interpretations of labor laws vary across statutes \cite{Bales_undated-ok, harris2015proposal}.

\subsubsection{Income Instability}
Gig workers also suffer from a lack of financial stability induced by job precarity and the temporary nature of contractual work \cite{anwar2020precarity}. In their study evaluating the job quality of gigs, Wood et. al. identified \add{how} algorithmic management of workers cause\add{s} financial instability, social isolation as well as overwork and exhaustion \cite{Wood2018-lq}. The combination of low pay, high job insecurity, long working hours induces a high sense of precarity among gig workers \cite{Dubal2017-bj,Hua2018-qx,Sutherland2020-wk,Webster2016-gi,Vosko2006-ec}. One major contributor to the income instability of gig workers is seasonality, endangering the financial security of part-time gig workers. For instance, work in sports has always been characterized as precarious and seasonal, and the suspension of several major sports during the pandemic has intensified such impacts \cite{javits2022gig, sheptak2020sport}. Ravenelle et al. also identified increased vulnerabilities of gig workers during the pandemic, finding knowledge, sociological, and temporal/financial hurdles that prevent their access to unemployment assistance \cite{ravenelle2021side}.

\subsubsection{Minimal Access to Working Necessities}
The growing prevalence of gig work probes at previously unexplored social barriers, highlighting inadequacies in our public infrastructure. In New York City, exploitative labor practices induced by platforms and public infrastructure subject food couriers to dangerous working conditions, leading to a local labor union of cyclists in 2019 -- \textit{Los Deliveristas Unidos} \cite{geschwindt2022biking}. Based on the lived experiences of its constituent deliveristas, the grassroots collective formed a list of five demands surrounding working conditions, including a right to 1.) free public bathroom access 2.) physical public space for eating, resting and protection from harsh weather conditions 3.) hazard pay for work performed that involve physical hardships (e.g. the COVID-19 pandemic) and 4.) protections from e-bike robberies, wage theft and health and safety hazards. While the city council passed a bill last year to ensure bathroom access for workers \cite{nycbill}, enforcement is difficult and deliveristas still report instances of restaurants who restrict bathroom access  \cite{bathroomreport}.

\subsubsection{Safety Concerns}
Without proper employment classification, gig workers do not enjoy the regulated safety assets provided to traditional workers (e.g. worker's compensation, health insurance, and unemployment insurance, among other laws and regulations) \cite{kost2020boundaryless, abraham2017measuring, kuhn2021human}. Unfortunately, the non-standard nature of many gig work arrangements raises occupational health and safety risks, increasing scholarly, legal, and societal concern \cite{orr2022necrocapitalism, Howard2017-wd, Almoqbel2019-in}. For instance, Ferrie et al. found that poor mental health outcomes can result from sudden unemployment \cite{Ferrie1998-th}, and by 2006, Virtanen et al.'s review of 27 case studies revealed a solid association between temporary employment and morbidity \cite{Virtanen2005-jp}. Over the past five years, the Markup has tracked a total of 361 ride-hail and delivery drivers as victims of carjackings or attempted carjackings \cite{Kerr_undated-ny}.

Underlying drivers' safety are factors that disincentivize them to self-protect. Almoqbel and Wohn uncovered that platforms' rating systems to prevent drivers from engaging in protective behaviors (e.g. using dash cams) due to passengers' discomfort around monitoring (which lead to poor reviews); they further found drivers to share safety resources, vent about passengers, and coordinate informal union activities in online forums \cite{Almoqbel2019-in}. Beyond physical attacks, Bajwa et. al. discussed how precarity, occupational and platform-based vulnerabilities can cause psychological distress, increased risk for traffic accidents and musculoskeletal injuries, as well as work-induced stress, respectively \cite{Bajwa2018-gy}. From the  perspective of international law, Howard discussed how legal misclassifications cause a loss of protections and benefits for workers across the globe \cite{Howard2017-wd}.

\subsubsection{Missing Collective Action Power}
The design and structure of online labor platforms creates unique challenges such as information asymmetries and power imbalances between workers and clients, giving rise to the platformic control and algorithmic management \cite{Lampinen2018-sv,Jarrahi2019-if, Sutherland2020-wk, Jarrahi2020-zz,Gray2019-ue, Newlands_undated-sl, Rosenblat_undated-gm}. 
Such dynamics disincentivize workers from engaging in collectivism due to fears of losing competitive advantages \cite{Yao2021-eb}. Furthermore, the lack of physical workspaces prevents workers from forming collectively identifies and protesting inequities \cite{chesta2019labour, calacci2022organizing}, while antitrust and employment laws legally prevent them from performing such collective actions \cite{donovan2016congress, paul2019fissuring}. It is also notable to mention that migrant workers comprise a growing portion of the platform labor market, but legal restrictions make it difficult for them to engage in union activities or benefit from national welfare systems \cite{van2020migration}.

To acquire more workplace gains and protections, workers can engage in collective labor activities. But as Yao et. al. and Johnston et. al. find, barriers such as geographic dispersal, individualistic nature of gig work, and platforms' opposition to worker organization, all prevent the building of a collective, group agency \cite{Yao2021-eb, Johnston_undated-wy}. Furthermore, ``antiquated notions of collective bargaining \dots surrounding the gig economy'' may not prove useful in the modern digital workforce \cite{Johnston_undated-wy}. Meanwhile, Khovanskaya et. al. leveraged historical insights from mid-20th century labor unions toward management to inform how contemporary data-driven worker advocacy can bring workers together over shared concerns and raise public awareness of working conditions, instead of engaging in bureaucratic negotiations with platforms \cite{Khovanskaya2019-xo}. But as Graham et. al. points out, there is a dearth of counterhegemonic research efforts particular to the gig economy that support the ``building of alternatives, outrage, conflict, and worker organization'', a gap that we hope to help fill \cite{graham2018towards, outsidetheboss}.

\subsection{Design Efforts to Study Worker Well-being}
Early efforts to combat algorithmic management arose in contexts of crowdwork (Amazon Mechanical Turk), rideshare driving, and food couriering. The pioneering piece along this line of work centered Turkopticon, a widely-adopted browser plug-in that overlays its requester/employer-reviewing features on top of the AMT site to resist minimal wages, low quality work, and unfair job rejections (a.k.a. wage theft). In the author's own words, the system aimed to ``make questions of work conditions visible among technologists, policy makers, and the media'' \cite{irani2013turkopticon}. A companion tool Dynamo was developed subsequently to facilitate collective organization action among AMT workers \cite{salehi2015we}. A ``social sensing'' probe developed by You et. al. collected and shared personal health data of rideshare drivers with their significant others to promote well-being awareness (especially related to long working hours) and motivate behavioral changes \cite{you2021go}. Zhang et. al. leveraged algorithmic imaginaries to expand participants' current understandings of algorithms so as to generate alternative futures that actually support workers' needs \cite{zhang2022algmanagement}. In \cite{bates2021lessons}, Bates et. al. hosted two rounds of co-design workshops with gig cycle couriers in the U.K. to identify challenges in their working conditions and ideate alternative solutions. Codesign has also been used to unearth the accounts of essential workers such as airport janitorial staff \cite{kang2022stories}. Finally, Alvarez de la Vega, et al. used design fiction (informed by prior literature) in focus groups to discover potential design opportunities for improving the well-being of online freelancers \cite{alvarez2022design}.

These studies all took a worker-centered focus to empower and highlight the voices of underserved workers. We expand beyond workers to capture the opinions of three distinct but relevant stakeholder groups, so that these involved parties may take part in constructing a brighter and improved gig work future. In particular, we hope our findings help policymakers make well-informed decisions when establishing new regulations to protect worker rights, as well as the media and public at large to exert pressure on platforms to implement worker-centered changes, benefits and programs.


\subsection{Multi-Stakeholder \& Solution-Centered Approach}


While the challenges that gig workers face are well-studied, few investigations have taken a holistic perspective to examine how adjacent stakeholders such as platform-side designers or policymakers can play a role in alleviating such constraints. 
By asking our participants to generate and rank solutions to these issues, we aimed to identify the most desired and practical improvements for addressing the challenges present in gig work (RQ2). As Howard identified in their commentary, the key question of who should be held responsible for providing various job protections has yet to be answered \cite{Howard2017-wd}, so we directly asked stakeholders about who should bring forth change (\ref{procedures}) and probed their solution rankings with follow-up questions surrounding underlying incentives and constraints (RQ1).
By eliciting such preferences and limitations, our workshops goes beyond worker perspectives to also explore unmet needs of platforms and policymakers, so as to help maximize their ability to support gig workers as advocates. 
Sociologists identified these three groups as key stakeholders of the gig economy \cite{vallas2020platforms}, and our simultaneous engagement with all three ensures that the solutions arising from our workshops are acceptable to and welcomed by multiple involved parties. 
In particular, we encouraged participates to generate their own solutions as a means of negotiating for potential futures that they find the most suitable. 
After all, many factors that harm worker well-being (e.g. legal misclassifcation, algorithmic management) can only be mitigated with solutions at systemic as well as cultural levels, and such changes require the active collaboration and involvement of lawmakers, platform designers, gig workers, as well as the public at large.

\section{Methods}
\begin{table*}[t]
\caption{Workshop IDs \& Participant Summaries}
\begin{tabular}{|c|l|c|p{7cm}|}
\hline
\textbf{Workshop ID} & \textbf{Stakeholder Group} & \multicolumn{1}{l|}{\textbf{\# Participants}} & \textbf{Relevant experience} \\ \hline
R1 & Regulators/Advocates & 3 & Manager at DHS; Director of community management at National Council of Jewish Women; intern analyst to director; \\ \hline
P1 & Platform employees & 2 & Executive recruiter at a major rideshare organization; Product designer and an ex-employee of multiple e-commerce platforms \\ \hline
W1 & Gig workers & 3 & 1 deliverer and 1 driver for a popular food delivery platform; nurse at a healthcare company; \\ \hline
R2 & Regulators/Advocates & 2 & Director of Mobility Dept for local city; Professor in organizational behavior and public policy \\ \hline
W2 & Gig workers & 5 & Full time food courier of 1.5 years; freelancer at a platform for matching local labor to demand; IT freelancer \\ \hline
R3 & Regulators & 2 & Local councilperson; Professor of Cyber Law, Policy, and Security \\ \hline
P2 & Platform employees & 2 & Product manager at a platform for matching local labor to demand; Program lead at a rideshare platform \\ \hline
P3 & Platform employee & 1 & Employee at a popular food delivery platform \\ \hline
\end{tabular}
  \label{tab:participants}
  \Description[Participant Summaries]{The table lists the individual stakeholders that belong to each workshop (ID's are listed in a column for each). Each stakeholders'™ unique experiences are summarized and workshops are labeled by their respective stakeholder groups.}
\end{table*}

\subsection{Recruitment and Participants} \label{recruit}
Our participant pool consisted of three stakeholder groups: gig workers, local regulators and members of various public service organizations, as well as employees from popular gig work platforms, who were chosen because they represent the groups that can actively become involved in solutions for improving gig worker well-being, independently or collaboratively. Gig workers can develop and practice their own strategies, policy-makers can enact laws to restrict  how platforms affect workers, platform employees can modify features to improve gig worker well-being, and together they can drive forth systemic changes that bring us closer to healthy and productive gig communities.

We recruited a total of 20 unique participants across 8 workshops. The seven participants from the regulator/advocates group were reached through contacts from the Pittsburgh-based research institute Metro21, and consisted of individuals who self-identified as regulators or worker advocates from local organizations such as the Department of Human Services and United Way. While not all of our regulator participants are actively involved in policy-making (some study public policy while others work for government agencies), we did recruit one councilperson. The eight gig workers responded to our recruitment posts on Reddit and included individuals who made earnings on popular ridesharing or food delivery apps. The last group consisted of five platform employees (e.g. product designers, managers, and engineers) whom we contacted through a combination of Reddit posts and LinkedIn direct messages. Participants selection was based on responses to a pre-screen survey, which asked for affiliated organizations and engagement with gig work(ers). Table \ref{tab:participants} summarizes the workshop participants and their relevant expertise, in chronological order of workshops dates.
\subsection{Study Design} \label{procedures}
\subsubsection{Speed Dating}
As the nature of gig work probes at previously unexplored social boundaries (e.g. traditional workers typically do not bear responsibility for consumers' physical or food safety), we require alternative methods for examining workers' needs, as well as to discover the social and cultural barriers that gig work pushes at, which are not yet well understood \cite{Davidoff2007-fq,Zimmerman2017-rq}.
Toward this end, we leveraged speed dating, a method that involved presenting pressing issues (design opportunities) and provocative alternative futures (design concepts) to multiple stakeholders in rapid sequence, enabling us to uncover their latent needs, desires, fears and dreams. 
Unlike romantic speed dating, where the goal is to pair potential couples, the technique strives to match gig work issues to potential solutions. 
Speed dating has been utilized in a variety of domains (e.g. attention management \cite{chou2022because}, AI ethics checklists \cite{madaio2020co} smart homes \cite{jin2022exploring}) to rapidly explore of concepts/solutions to issues without needing to implement the proposed technologies \cite{Davidoff2007-fq}. 

Most similar to our contexts, Dillahunt et. al. found speed dating effective in identifying concepts for addressing needs of underserved job seekers \cite{Dillahunt2018-ia}. 
Following their study design, we presented to participants a series of issues that gig workers face, but did not pair each issue with a tool/design concept in the same way. 
Instead, we offered a list of alternative futures (and encouraged participants to generate their own solutions) to broaden the horizon of imagined possibilities. 
While parts of our study design drew inspiration from \cite{Dillahunt2018-ia}, we center our work around gig workers instead of underserved job seekers, and expand the pool of imagined solutions by incorporating the voices of diverse stakeholder groups. 

\subsubsection{Scenario Construction}
Initially, we generated ten scenario stories and subsequently solicited the critique of other researchers working in the space of supporting gig workers to help us finalize a problem space comprising five scenarios (see Table \ref{tab:scenarios}).
The scenarios were developed based on challenges outlined in relevant literature as well as pressing issues that received press coverage. In particular, the fourth \cite{Al_Jazeera2022-vz} and fifth \cite{noauthor_undated-xr} scenarios were conceived based on accounts of stories of worker situations covered in the respective articles. Each scenario maps back to the respective subsubsection in 
\ref{challenges}.
To avoid promoting ``blue-sky'' thinking, which (as Harrington et. al. pointed out \cite{harrington2019deconstructing}) may lead to frustration for the very population we intend to serve, the authors collectively generated ideas ahead of time to prepopulate the solution space (which consisted of ideas implementable by each of the three involved stakeholder groups to avoid imbuing our opinions on who should hold responsibility), so as to help participants brainstorm -- a full list of pre-generated solutions can be found in Table \ref{tab:solutions} of the Appendix. 

Though all scenario characters were fictitious, the first three were inspired by concerns expressed during a local workshop organized by the National Council of Jewish Women, which explored the hidden costs of gig work. The fourth \cite{Al_Jazeera2022-vz} and fifth \cite{noauthor_undated-xr} scenarios were based on accounts of stories of worker situations covered in the respective articles. All five scenarios represent prevalent issues gig workers face today: missing employee benefits, financial instability, a lack of essential working necessities, safety issues and workers' minimized ability to take collective action. With the exception of the persona in Scenario 3, who reflects the common characteristics of food deliverers (i.e. male, young, and of an immigrant background \cite{ma2022brush}), the demographics of characters are intentionally non-representative of the general gig worker population to encourage the consideration of marginalized workers (women, elders, etc.), who often face issues such as bias, harassment, and pay gaps, all of which intersect with algorithmic control \cite{ma2022brush, anjali2021watched, foong2021understanding, foong2018women, jahanbakhsh2020experimental}.

\subsubsection{Storyboards} To present these scenarios, we constructed five pictorial storyboards depicting stories based on news articles, local workshops, and prior work. Storyboarding, defined as ``a short graphical depiction of a narrative'', is an effective tool for demonstrating 1.) impacts of technologies on human activity and 2.) effects of proposed (technological) interventions and solutions before implementation. Since we cover a wide range of gig worker types in this study (e.g. food couriers, rideshare drivers, movers and online sellers), storyboards allow participants to quickly engage with specific situations, connecting their own lived experiences when applicable. 
Following  Truong et. al.'s guidelines \cite{truong2006storyboarding} on best practices for storyboarding (concise background, intentional text, characters, graphics, passing of time, etc.), we drew empathy from our participants using personas of gig workers, included text to orient participants in the character's world, and only constructed three frames per scenario to succinctly convey each character's activities to avert bogging participants down with overt details.

\subsubsection{Procedures}
Each scenario was presented via three storyboard cards, and we guided conversation using a probing question that focuses discussions around broader underlying issues. 
After introducing the scenario and probing question, we requested that participants read the prepopulated solutions and treat them as seed solutions for generating their own ideas, and subsequently \textbf{rank all the solutions for the scenario}. Table \ref{tab:solutions} and \ref{tab:rankings} in the Appendix show the generated solutions and an example instance of the ranking process. 
During the ranking process, we solicited the rationales of participants' ranking decisions to probe at and uncover latent social boundaries and desiderata. Due to time constraints, we did not engage our participants in a formal consensus building processes (e.g. the Delphi method) during rankings. After solution ranking, we asked a set of followup questions to wrap up each scenario. The scenarios were presented in the same order across all workshop sessions, as shown in Table \ref{tab:scenarios}.

After completing the above, participants were asked to \textbf{rank the five scenarios} in terms of what they thought were most important to address, effectively performing needs-validation over the issues we presented. In summary, we asked participants of each workshop to complete the following set of tasks, in order: 
\begin{enumerate}
    \item \add{For each of the five scenarios:}
    \begin{enumerate}
        \item \add{Examine the scenario's storyboard and accompanying descriptive text (including the probing question)}
        \item Read through and discuss the list of prepared solutions, then add newly generated ideas
        \item \add{Rank the solutions (including the ideas generated live) based on preferences and priorities, using sticky notes}
        \item \add{Explain reasoning for ranking preferences}
        \item \add{List the most and least preferred solutions}
        \item \add{Express who should be responsible for implementing the mentioned solutions (using provided check-boxes)}
    \end{enumerate}
    \item \add{Rank the five scenarios in terms of which issues are most important to address}
\end{enumerate}
Participants were encouraged to add solutions at any point in these steps. Additional materials used for workshops are included in supplementary materials, and solutions generated by participants are available in the Appendix.

\begin{table*}[t]
  \centering
   \caption{Problem Space: Scenario Summaries}
    \includegraphics[width=0.875\textwidth]{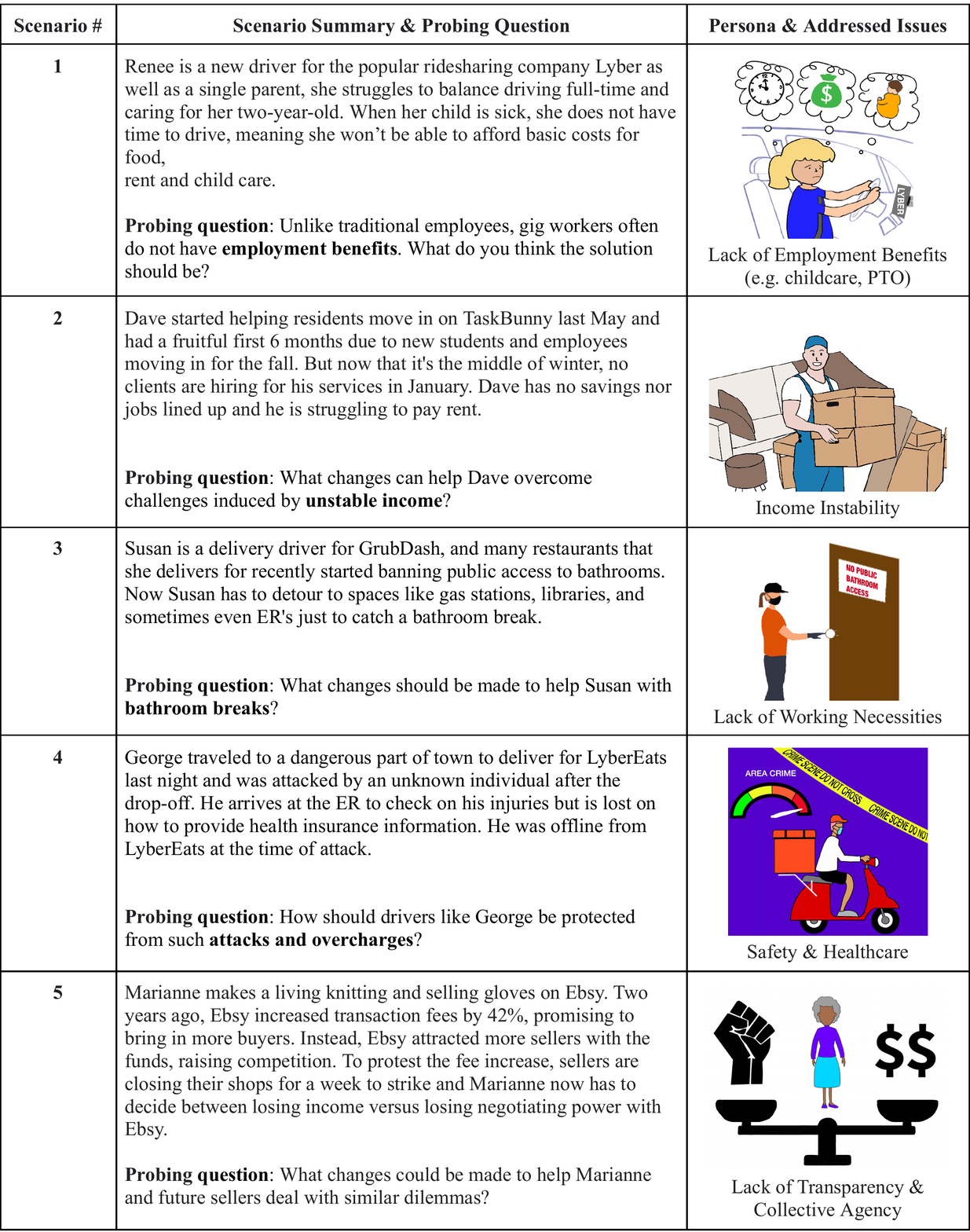}
\Description[A table summarizing five scenarios]{For each scenario, the table presents a summary of the persona's dilemma, as well as a thumbnail depicting their situations along with the higher-level issues that the scenario points to.}
  \label{tab:scenarios}
\end{table*}

\subsection{Workshop Setup}
We conducted a total of 8 co-design workshops with 20 participants, one of which was in-person while the rest were virtually conducted via Zoom. All participants were located in the United States and compensated at a rate of \$60/hour for their time, and each workshop lasted 90-120 minutes. To encourage discussion and collaboration among participants of the same stakeholder group, we included 2-3 participants in most workshops instead of conducting individual sessions. 
Combining the gig workers with the policy makers or platform employees could have discouraged workers to speak up in workshops, and thus we only included one stakeholder group in each workshop (Table \ref{tab:participants} indicates the relevant stakeholder group to each workshop). This separation was intended to avoid further disempowerment of already marginalized voices, and to minimize the emergence of power differentials that could have resulted from potential employment relationships  -- workers in one group may have been demotivated to express their honest opinions if the workshop also hosted their employer.
Because we studied our stakeholder groups separately, participants were able to connect and collaborate easily with peers from similar backgrounds. This setup of groups with similar experiences and values made each co-design workshop a productive discussion rather than confrontational. We also helped different participant groups collaborate asynchronously with each other by updating them on relevant solutions and rankings from previous workshop sessions.

Prior to each workshop, we set up whiteboards on Miro or physical easel pads to present the scenarios and potential solutions to participants, which served as a space for participants to rank or add solutions via sticky notes, and to document their finalized preferences. We took video recordings and field notes across workshops and collected participants' solution rankings, votes on who should take responsibility, and newly generated solutions.

\subsection{Positionality}

As Irani states, reflexivity in HCI allows us as researchers to produce better knowledge by ``recognizing designers' positions, values, limitations, and standpoints''. In the following, we reflect on our own positions as designers and researchers as well as how it impacts our work outputs \cite{irani2016stories}. We are all researchers residing in the US who work or receive training in the fields of Computer Science, Human Computer Interaction, and Music. Two of us live in the city where the study was conducted and have prior experience conducting research surrounding gig work. However, we recognize the relative privileges we hold in society as compared to worker participants. For instance, none of us have completed gig work ourselves and therefore lack first-hand experience of the issues that gig workers face. Additionally, we have all been a consumer on gig platforms; three authors often speak to rideshare workers about their job during rides and one author has family members who engage in gig work. The funding for this research sources solely from the National Science Foundation, and the work is not sponsored by any external companies or platforms.

We as researchers all hold the view that the current state of the gig economy, as discussed in Section 2, is incapable of supporting the well-being of gig workers and that these challenges should be addressed soon since it seems that gig work is here to stay. To address the need for change, we employ a combination of transformative, postmodern and pragmatic frameworks to interpret and understand the present day conditions of gig work, as well as to find practical approaches toward addressing some of these real world issues. \cite{Creswell2016-rq}. We presented day-to-day scenarios of individuals, which inform design decisions for addressing issues of gig work, and allows participants to generate solutions. In addition, we pre-populated the solution space with provocative ideas so as to give participants space for imagining more systematic solutions, which can contribute toward long-term reform of the gig-economy. 

Following best practices suggested by prior literature \cite{reinharz1992feminist, calvin, liang2021embracing}, we shared our affiliations and intentions with participants prior to workshops, reflected on our own biases as researchers, and pondered ``how can participants benefit from the study beyond the monetary compensation?'', ``are we bringing positive impacts to the worker community?'' and ``how can we place workers' ideas in a larger field of power?''.
We also include in \ref{limits} participant reflections on our study and considerations for future lines of research.

\begin{table*}[h]
\caption{Summary of Stakeholder's Motivations and Deterrents}
     \Description[Bulleted summaries of each stakeholder groups' incentives and deterrents]{Each stakeholder groups' distinct motivations, preferences and disincentives are outlined.}
\begin{tabular}{|l|l|l|}
\hline
\textbf{Stakeholders} & \textbf{Motivating factors and preferences} & \textbf{Deterrents} \\ \hline
Platform & \begin{tabular}[c]{@{}p{6.5cm}@{}} $\bullet$ Minimize worker decommission\\  $\bullet$ Required compliance to mandates and regulations \\  $\bullet$ Preserve public image\end{tabular} & \begin{tabular}[c]{@{}p{6cm}@{}}$\bullet$ Increased operation costs\\ $\bullet$ Thin profit margins \& market competition \\$\bullet$ Legal liabilities\end{tabular} \\ \hline
Workers & \begin{tabular}[c]{@{}l@{}}$\bullet$ Leverage multiple platforms\\ $\bullet$ Personalized solutions\end{tabular} & \begin{tabular}[c]{@{}p{6cm}@{}} $\bullet$ Disruptors to earning opportunities or client relations\\ $\bullet$ Short-term or unreliable solutions\end{tabular} \\ \hline
Regulators & \begin{tabular}[c]{@{}p{6.5cm}@{}}$\bullet$ Worker-initiated collective action\\  $\bullet$ Hold platforms responsible for initiating and implementing solutions that benefit their workers\end{tabular} & \begin{tabular}[c]{@{}p{6cm}@{}}$\bullet$ Providing special accommodations to specific worker subgroups \\ $\bullet$ Invasive monitoring of workers\end{tabular} \\ \hline
\end{tabular}
  \label{tab:stakeholders}
\end{table*}

\subsection{Analysis}
To begin analysis, we first computed average rankings for each solution and extracted the three highest and lowest ranked solutions for each scenario based on these averages. We then engaged in a thematic analysis approach to analyze 14 hours of Zoom recordings (transcribed by the online service \textit{Rev.com}) and 18 pages of field notes. In the first stage of the analysis, we followed an opening coding approach, where one to two researchers independently conducted qualitative coding for each workshop's data (at least one of these coders was present at the corresponding workshop) \cite{patton1990qualitative, mcdonald2019reliability, Patton2014-ef, corbin2014basics, strauss1990basics}. During this process, coders remained receptive and looked for as many codes as possible, while keeping in mind our research questions on worker well-being, the issues that each scenario targets, and potential future changes. The coders met to refine and resolve any disagreements about the initial codes, resulting in a total of 567 unique codes. In the next stage of analysis, we iteratively combined these codes into emergent themes and subthemes, wrote descriptive memos, and built an affinity diagram to map the relationships between categories \cite{holtzblatt2014contextual, Beyer1999-hr}. This analysis produced 8 themes and 63 subthemes, and we describe these findings below. The first set of findings
gives an overview of participants' rationales for rankings across scenarios, the second set reports on scenario-based themes from participant's reactions and perspectives on our proposed solutions, and the last set describes themes from participant-generated solutions.

\section{Results}
Each stakeholder group offered unique reactions to our scenarios and proposed solutions. Thus, we start by presenting overarching incentives and preferences that motivates each stakeholder group to initiate change, as well as factors that prevent them from implementing suggested solutions. Next we delve into individual scenarios to unfold participants' quantitative rankings of solutions and provide a debrief of their rationales using qualitative results. We end by describing participants' imagined solutions that spanned across workshops and scenarios.

\subsection{Multi-Stakeholders' Incentives, Preferences and Deterrents for Improving Gig Worker Well-being}

In this section, we present themes that emerged across various scenarios, reporting on stakeholders' overall incentives and preferences that motivate them to promote change for improving gig worker well-being, as well as factors that deter them from implementing suggested solutions. These patterns were revealed through discussions during solution-ranking/generation; Table \ref{tab:stakeholders} summarizes these findings. 

\subsubsection{Platform Motivations \& Preferences}
\paragraph{Minimize Worker Decommission}
Platforms are inherently incentivized to support participating workers, since their operations depend critically upon labor supply. 
For example, when workers are decommissioned, platforms are motivated to bring them back on a job because ``if the worker's not making money, if the worker's not available to work or just isn't working, the platform is not making money'' (P1). 
Worker decommission can result from a variety of factors, including fluctuating seasonal demands, a lack of opportunities or unmet childcare needs: ``If somebody doesn't have childcare, that does make them less likely to be available for work on the platform, which is problematic for the platform'' (P1). 

\paragraph{Government Mandates and Regulations}
Regulatory pressure can incentivize platforms to make changes, but an excess of mandates can cause them to ``think that a lot of this regulation stifles innovation'' (P1). Mandates are also undesirable to platforms because since they mean ``that we're more restricted, that we're gonna have to pay more'' (P1). In addition to restricting platforms from implementing novel features, the cost of (unfunded) mandates can also ``significantly restrict our bottom line and our ability to continue to function as a platform'' (P1).

\paragraph{Preserving Public Image}
To circumvent additional regulations, platforms are willing to implement services to preserve public image and ``appease the general public or regulators or media \dots by offering something like a childcare program'' (P1). Platforms' aversion to regulation is strong enough to dedicate ``large government relation teams that \dots strongly lobby against'' mandates ``except where they think that it benefits them to show the public for PR reasons'' (P1).

\subsubsection{Deterrents for Platforms}
\paragraph{High Operation Costs}
Many of the solutions we presented called for the development of services or programs benefiting workers. Platforms cited high costs and the prioritization of other services as reason against implementation:  ``if we're adding incremental benefits, we have to reduce something else'' (P1). According to one participant from P1,  implementing a single feature can cost ``easily six months of three engineers time, plus maybe a month of design effort, plus \dots you're probably talking about an initiative it's gonna cost \$650,000'', and such initiatives may be so ``prohibitively expensive, to the degree [that] the platform might not continue to be sustainable''.
\begin{table*}[h]
\caption{Scenario 1 Rankings and Voting Summary}
\begin{tabular}{|l|llc|}
\hline
\multirow{8}{*}{\begin{tabular}[c]{@{}c@{}}\includegraphics[width=2.75cm]{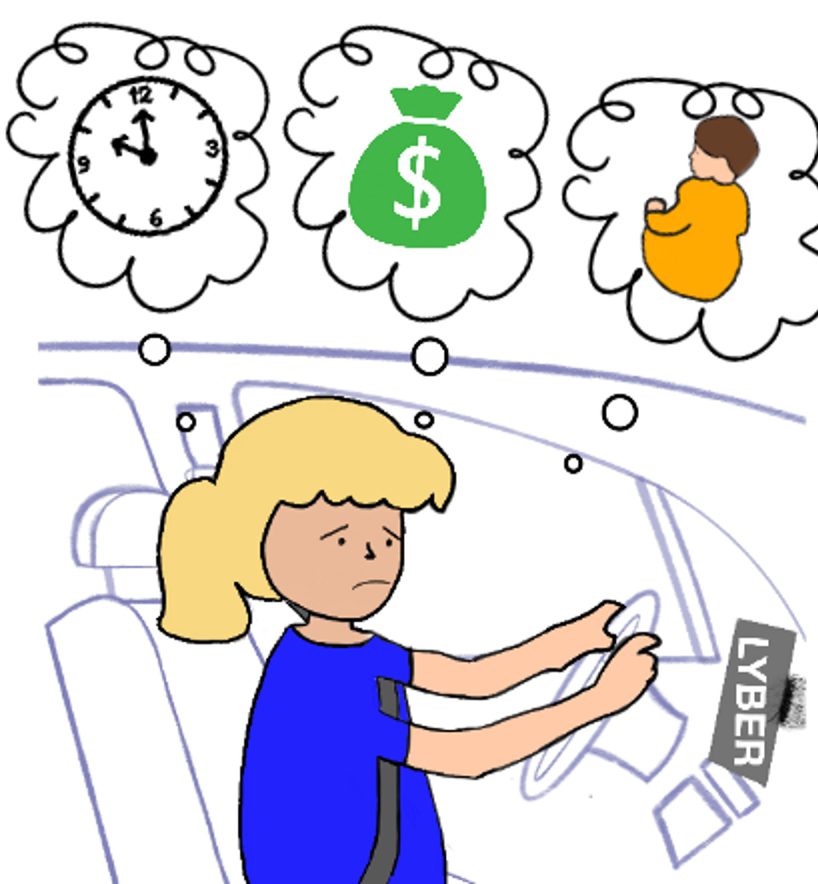} \\ \add{Renee balancing} \\ \add{rideshare work} \\ \add{and childcare.}\end{tabular}} & \multicolumn{2}{c|}{\textbf{Scenario 1} \add{(Lack of employment benefits)}}                                &  \begin{tabular}[c]{@{}c@{}}\textbf{Avg Ranking} \\ (lower = preferred)\end{tabular} \\ \cline{2-4} 
                  & \multicolumn{1}{l|}{\multirow{3}{*}{\begin{tabular}[c]{@{}c@{}}Top 3 \\ most \textbf{favored}\\  solutions\end{tabular} }} & \multicolumn{1}{l|}{{Platform offers childcare program {[}R1-2, W1-2, P3{]}}} & 2.625 \\ \cline{3-4} 
                  & \multicolumn{1}{l|}{}                  & \multicolumn{1}{l|}{Paid Time Off (PTO) {[}R1-3, W2{]}} & 3.313 \\ \cline{3-4} 
                  & \multicolumn{1}{l|}{}                  & \multicolumn{1}{l|}{Driver-support groups {[}R3, W1, P1{]}} &  3.313 \\ \cline{2-4} 
                  & \multicolumn{1}{l|}{\multirow{3}{*}{\begin{tabular}[c]{@{}c@{}}Top 3 \\ most \textbf{disliked}\\   solutions\end{tabular}}} & \multicolumn{1}{l|}{Platform offers higher hourly pay {[}R1, W1-2, P1-2{]}} & 5.250 \\ \cline{3-4} 
                  & \multicolumn{1}{l|}{}                  & \multicolumn{1}{l|}{\begin{tabular}[l]{@{}l@{}}Worker adds incentives to encourage tips \\ {[}R2-3, W1{]}\end{tabular}} & 4.625 \\ \cline{3-4} 
                  & \multicolumn{1}{l|}{}                  & \multicolumn{1}{l|}{\begin{tabular}[l]{@{}l@{}}Knowing the destination of incoming rides \\ {[}R1, R3, W1, P2{]}\end{tabular}}  & 4.688 \\ \cline{2-4} 
                  & \multicolumn{1}{l|}{\begin{tabular}[c]{@{}c@{}}Who should be \\ responsible for \\ making changes\end{tabular} }                  & \multicolumn{2}{l|}{\begin{tabular}[c]{@{}l@{}}$\bullet$ 7 of 8 workshops voted platforms [R1, R2, R3, W1, W2, P2, P3] \\ $\bullet$ 4 of 8 workshops voted workers [W1, W2, P1, P2] \\ $\bullet$ 5 of 8 workshops voted regulators [R1, R2, R3, P2, P3] \end{tabular}}    \\ \hline
\end{tabular}
  \label{tab:s1}
  \Description[A summary of stakeholders' responses to solutions presented for scenario 1]{The top three most and least popular solutions for scenario one are listed along with their workshop ID's. In this scenario Renee struggles to balance time shared between ridesharing and childcare. The (reverse) ranking averages across workshops are also offered to elucidate the strength of preference. Participants' votes on who should be responsible for changes is tallied.}
\end{table*}
\paragraph{Thin Profit Margins}
One might suggest that platforms use resources gleaned from profit margins to develop features that promote worker well-being. However, platform-side participants relates how ``margins are getting tougher and tougher on a lot of these products and services'' (P1). In order to provide for increased pay or benefits, ``the platform effectively needs to take less'', but ``the company's not really gonna take less cut because [then] they couldn't pay their employees and they just have to cut heads'' (P1). Alternatively, platforms can ``increase price [of its service]'', but that instigates a negative cycle by putting the platform at risk for user abandonment because if ``you raise it too high, you lose customers automatically, they don't wanna pay 50 bucks to go five miles'', so it ``reduces the number of users that will use the platform, which will cause Lyber to make less'' (P1).

\paragraph{Competition Between Platforms}
Exacerbating monetary constraints, customers were deemed ``very price sensitive, they're fickle, they may open both [apps]'' (P1). If they are not satisfied with prices, clients might just abandon the service altogether: ``There is a maximum amount of money that Lyber passengers are willing to pay for a single trip where [they] start to see declines in usage'' (P1). In fact, platforms assign ``an entire revenue optimization team that figures out how much can be charged and how much people are willing to pay.'' (P1).

\paragraph{Legal Liabilities} 
In addition to costs, another factor that demotivates platforms from service offerings is their potential legal ramifications. Platform participants fear such potential complications and ``hope that there wouldn't be reputational risk to Lyber by Renee's[/workers'] kid[s], potentially getting injured by being taken care of by another parent'' (P1). Regulator participants also recognized the risks, noting that ``one of the reasons why childcare programs aren't on sites in corporations is [because] the liability is huge'' (R2). 
The ambiguous legal classification of gig workers also disincentives additional provisions of benefits since ``the more that you \dots treat somebody as if they're an employee, the more they can argue in court that they are an employee'' (P1). 


\subsubsection{Worker Practices, Motivations \& Preferences}
\paragraph{Leverage Multiple Platforms} \label{multiapp}
To address instability, workers related experiences of engaging with multiple platforms at once: ``if things slow down on one platform, then you can go to another'' (W2). Distributing worker profiles across multiple platforms raises opportunities of procuring gigs, and workers view the labor of finding work as their own responsibility: ``you can't just sit there and say that TaskBunny should be responsible \dots when it's off season, it's upon you now to maybe seek other alternatives of earning'' (W1).

\paragraph{Personalized Solutions} \label{personalized}
The instability of gigs often forces workers to fit needs around work schedules, but ironically the promised flexibility is oftentimes what drove them toward gigs in the first place \cite{lee2015working}. Thus it's on platforms to adjust around worker schedules, ``to understand the kind of situation that you're in and then they'll try to adjust to fit your availability \dots this is the best way \dots [when] they're trying to adjust to your schedule \dots [and] to your situation'' (W1). Adjusting to workers' circumstances can provide a peace of mind through both regularity on standard days and accommodations during emergencies. Platforms don't currently account for situations where ``[there is an] employee who is on maternity leave \dots [or] away for stuff like funerals'', but workers desire solutions that consider ``the various kinds of condition[s] that needs them to be away from work'' (W1).

\subsubsection{Deterrents for Workers}
\paragraph{Impediments to Earning or Damages to Client Relations}
Worker participants held a strong aversion against changes that conflict with their own priorities (e.g. making earnings, maintaining good reputation with clients). For example, when presented with Susan's predicament of being blocked from restaurant bathrooms, one worker explained how ``you need to work to get money'', challenging the hypothetical idea that if ``all the restaurants fail to offer bathroom services, do you stop working?'' (W1). Another worker opposed ``the restriction of platforms, [since] that means you wouldn't have work'' (W2). They were also mindful of client relationships, stating concerns that ``avoid[ing] orders from those locations, meaning that the clients would suffer'' (W2)
. Beyond clients, workers also ``wouldn't want to get on a restaurant's bad side'' (W2).


\begin{table*}[]
\caption{Scenario 2 Rankings and Voting Summary}
\begin{tabular}{|l|llc|}
\hline
\multirow{8}{*}{\begin{tabular}[c]{@{}c@{}}\includegraphics[width=2.75cm]{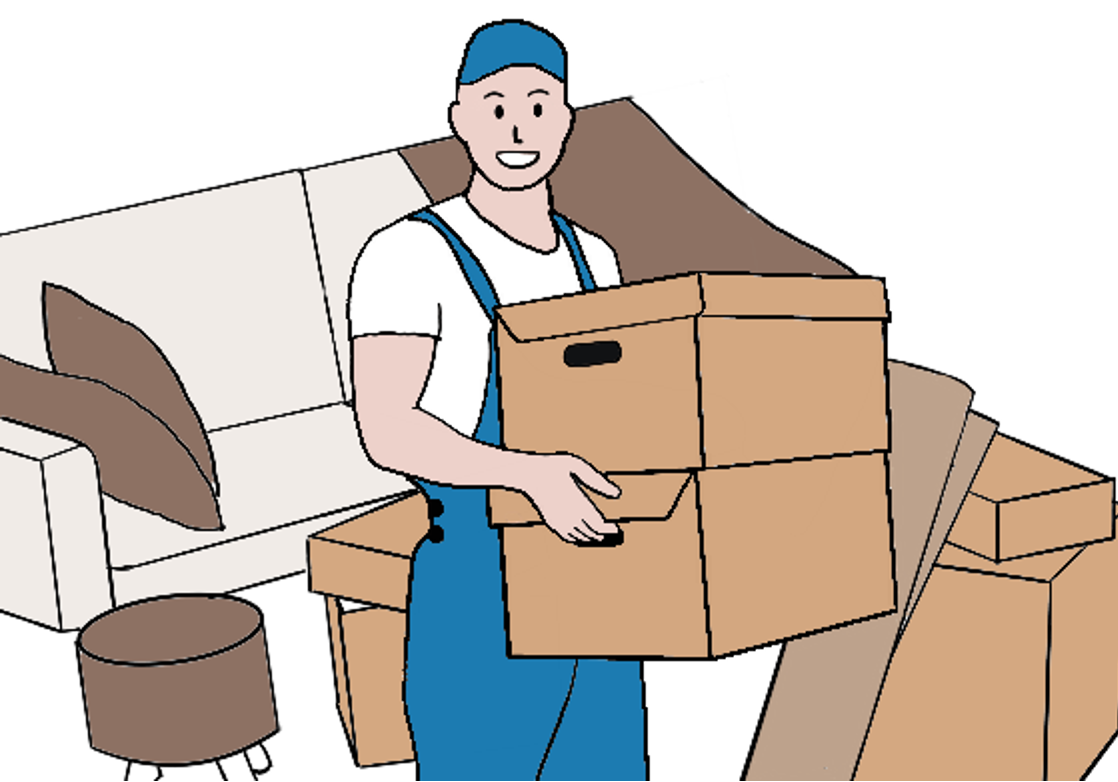} \\ \add{Dave facing}\\ \add{seasonal lows in} \\ \add{job opportunities.}\end{tabular}} & \multicolumn{2}{c|}{\textbf{Scenario 2} \add{(Income instability)}}                                & \textbf{Ranking} \\ \cline{2-4} 
& \multicolumn{1}{l|}{\multirow{3}{*}{\begin{tabular}[c]{@{}c@{}}Top 3 \\ most \textbf{favored}\\  solutions\end{tabular} }} & \multicolumn{1}{l|}{Winter side hustles/off-season work recommendation {[}all{]}} & 1.000 \\ \cline{3-4} 
& \multicolumn{1}{l|}{}                  & \multicolumn{1}{l|}{Platforms plan events during off seasons {[}R2, P1, P3{]}} & 3.875 \\ \cline{3-4} 
& \multicolumn{1}{l|}{}                  & \multicolumn{1}{l|}{Workers conduct long-term financial planning {[}W1-2{]}} & 4.188 \\  \cline{2-4} 
& \multicolumn{1}{l|}{\multirow{3}{*}{\begin{tabular}[c]{@{}c@{}}Top 3 \\ most \textbf{disliked}\\   solutions\end{tabular}}} & \multicolumn{1}{l|}{Workers conduct long-term financial planning {[}R1-2, P1-2{]}} & 4.188 \\ \cline{3-4} 
& \multicolumn{1}{l|}{}                  & \multicolumn{1}{l|}{Platforms plan events during off seasons {[}R3, W1-2, P2{]}} & 3.875 \\ \cline{3-4} 
& \multicolumn{1}{l|}{}                  & \multicolumn{1}{l|}{Regulators provide unemployment benefits {[}R2, P1{]}} & 4.188 \\ \cline{2-4} 
& \multicolumn{1}{l|}{\begin{tabular}[c]{@{}c@{}}Who should be \\ responsible for \\ making changes\end{tabular} }                  & \multicolumn{2}{l|}{\begin{tabular}[c]{@{}l@{}}$\bullet$ 8 of 8 workshops voted platforms [R1, R2, R3, W1, W2, P1, P2, P3] \\ $\bullet$ 5 of 8 workshops voted workers [R1, W1, W2, P1, P3]\\ $\bullet$ 3 of 8 workshops voted regulators [R2, W2, P2]\end{tabular}}    \\ \hline
\end{tabular}
  \Description[A summary of stakeholders' responses to solutions presented for scenario 2]{The top three most and least popular solutions for scenario two are listed along with their workshop ID's. In this scenario Dave faces the seasonal lows when earning on TaskBunny. The (reverse) ranking averages across workshops are also offered to elucidate the strength of preference. Participants' votes on who should be responsible for changes is tallied. }
  \label{tab:s2}
\end{table*}
\paragraph{Short-term or Unreliable Solutions}
Temporary solutions were also undesirable to workers, as they offer only short-lived relief to long-lasting problems. While some help is better than nothing, ``they are just short term, they may be a day or two solutions in a month, in the whole season'' (W2). For childcare needs, ``[days of paid time off] is not a solution because \dots she has to stay with the kid'' (W1). Worker participants also resisted solutions out of their control, since they may be breakable -- ``security equipment could fail, maybe the cameras have failed to work, or failed to capture a clear image of the attack'' (W1) -- or unreliable -- ``off-season events that are planned by TaskBunny maybe would not be very reliable'' (W2).

\subsubsection{Regulator Motivations \& Preferences}
\paragraph{Hold Platforms Accountable}
Regulator participants held companies largely responsible to creating better working conditions for their employees. One R3 participant emphasizes how ``it's the company's responsibility to create a work environment that is conducive to people succeeding and building the lives that they want''. Specifically, they ``could imagine a world in which the platform invests in safe bathroom facilities for their own people'' (R3). In addition to bathroom access, one regulator also contended that ``platforms are viable for healthcare consequences associated with the work that their people are doing'' (R3).

\paragraph{Worker-initiated Collective Efforts} \label{worker_action}
Power and informational asymmetries makes regulators ``reluctant to say the burden should fall on one person's shoulder to save themselves'' (R3). Instead, regulator participants recommended ``finding ways for the gig workers to combine effectively'' (R3), through collective worker actions such as pooling, unionizing and striking to impose pressure on platforms to initiate change. 
But since gig workers are not employees, many questions exist around how to collectively organize and bargain: ``How do you strike when you're not a union? How do you strike and what do you demand?'' (R2). 
Soliciting company involvement was one potential solution: ``If not in a formal union, having a company that gives their employees the opportunity to convene and to say what matters most to them could be good as a company practice or policy'' (R2).

\subsubsection{Deterrents for Regulators}
\paragraph{Special Accommodations for Particular Subgroups}
Regulators repeatedly emphasized inclusion (of workers and customers alike) and resisted special accommodations for specific groups. They raised additional questions like ``Do you have it for the single dad? Do you have it for like elder care? Where do you stop?'' (R3). 
For instance, while the idea of issuing badges to workers helps with limits on bathroom access, it also prompts problems of privacy and misuse: ``thing about badges \dots is that even if they're voluntary, any program of self-identification creates risks \dots with prospective privacy vulnerable populations, you can't really predict how that kind of information is going to circulate and be used in an inappropriate way'' (R3). 
In general, regulator participants objected to ``the idea of demarcating workers differently \dots that's dangerous and creates fault lines between people \dots even if \dots you're not closely tied to each other'' (R2). Thus, it's imperative ``for the company to have its own policies (designed either by mandate or by voluntary corporate structure) to be as inclusive [of] as many different types of workers as possible'' (R3). 

\paragraph{Violations of Worker Privacy}
Regulators also opposed invasive monitoring of workers, citing a violation of basic human rights. 
For example, ``a single mom badge come with risk \dots [you can imagine] some sketchy dude who likes to pick up women with kids and abuse them, then I think identifying someone as such could lead to safety concerns'' (R2). 
Another participant protests how ``we've gotten to the point where, because of technology and oversight, people have literally no independence - they can't even go to the bathroom on their own [initiative] anymore \dots [it's] kind of a human rights violation to have that kind of deep oversight of your employee'' (R3).
Monitoring via dash cams also pose issues of invasion, for while they allow workers ``to share [footage]\dots with the police so that they can help solve the crime'', they may also be ``pointing in at them as they're driving, I could see just a huge amount of privacy concerns rising from that'' (R3).

\subsection{Scenario Rankings and Rationales}
In the following scenario-based analysis, we include the top three most favored solutions as well as disliked solutions, and indicate the workshops that casted their votes via a bracked list of workshop IDs. Some solutions triggered polarizing opinions across different stakeholder groups, and may therefore simultaneously appear as both favored and disliked. To elucidate the strength of preference, we include the average rankings of individual solutions across all workshops, where lower rankings indicate more preferred solutions. To summarize each scenario, we wrap up with a recap of tensions between stakeholder groups and acceptable solutions that are common grounds to multiple stakeholder groups.

\begin{table*}[]
\caption{Scenario 3 Rankings and Voting Summary}
\begin{tabular}{|l|llc|}
\hline
\multirow{8}{*}{\begin{tabular}[c]{@{}c@{}}\includegraphics[width=1.5cm]{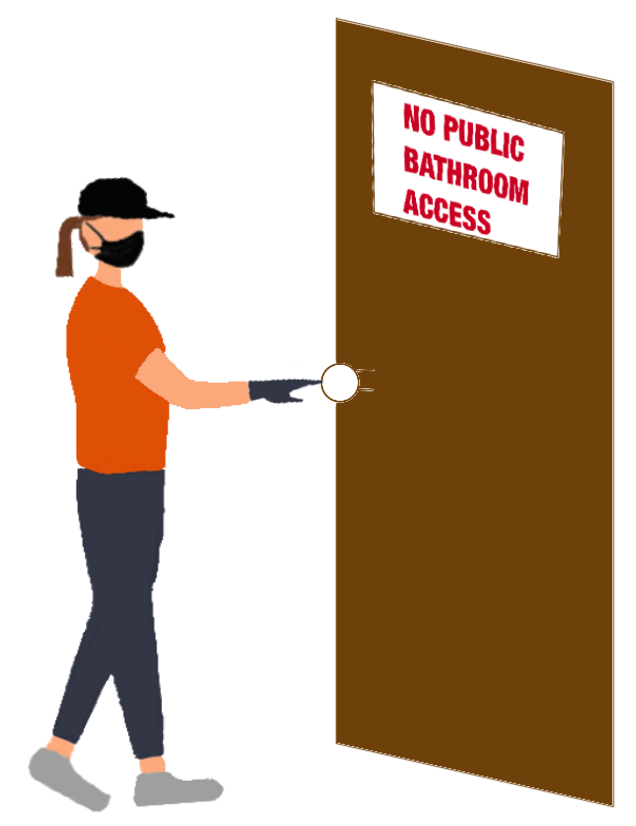} \\ \add{Susan struggling}\\ \add{to access bathrooms} \\ \add{at restaurants that} \\ \add{she delivered for.}\end{tabular}} & \multicolumn{2}{c|}{\textbf{Scenario 3} \add{(Missing Access to Working Necessities)}}                                & \textbf{Ranking} \\ \cline{2-4} 
& \multicolumn{1}{l|}{\multirow{3}{*}{\begin{tabular}[l]{@{}c@{}}Top 3 \\ most \textbf{favored}\\  solutions\end{tabular} }} & \multicolumn{1}{l|}{\begin{tabular}[l]{@{}l@{}}Platforms negotiate with restaurants \\ to open bathroom locations to workers. {[}R1, W1-2, P2-3{]}\end{tabular} } & 2.188 \\ \cline{3-4} 
& \multicolumn{1}{l|}{}                  & \multicolumn{1}{l|}{\begin{tabular}[l]{@{}l@{}}
Platforms show public bathroom locations in apps. \\ {[}R1, P1-3{]} \end{tabular}} & 2.125 \\ \cline{3-4} 
& \multicolumn{1}{l|}{}                  & \multicolumn{1}{l|}{\begin{tabular}[l]{@{}l@{}}Regulators require restaurants to provide bathroom access. \\ {[}R1-2, W1-2{]}\end{tabular}} & 4.188 \\  \cline{2-4} 
& \multicolumn{1}{l|}{\multirow{3}{*}{\begin{tabular}[c]{@{}c@{}}Top 3 \\ most \textbf{disliked}\\   solutions\end{tabular}}} & \multicolumn{1}{l|}{\begin{tabular}[l]{@{}l@{}}
Platforms cut off online orders during busy hours. \\ {[}R1-3, W1, P1-3{]}
\end{tabular}} & 7.688 \\ \cline{3-4} 
& \multicolumn{1}{l|}{}                  & \multicolumn{1}{l|}{Workers petition restaurants for bathroom access. {[}R2, P2{]}} & 5.375 \\ \cline{3-4} 
& \multicolumn{1}{l|}{}                  & \multicolumn{1}{l|}{\begin{tabular}[l]{@{}l@{}}
Workers share public bathroom locations with \\ one another.  {[}W1, P2{]}\end{tabular}}  & 5.188 \\ \cline{2-4} 
& \multicolumn{1}{l|}{\begin{tabular}[c]{@{}c@{}}Who should be \\ responsible for \\ making changes\end{tabular} }                  & \multicolumn{2}{l|}{\begin{tabular}[c]{@{}l@{}}$\bullet$ 8 of 8 workshops voted platforms [R1, R2, R3, W1, W2, P1, P2, P3]\\ $\bullet$ 3 of 8 workshops voted workers [W1, W2, P1]\\ $\bullet$ 7 of 8 workshops voted regulators [R1, R2, R3, W1, W2, P1, P2]\end{tabular}}    \\ \hline
\end{tabular}
  \Description[A summary of stakeholders' responses to solutions presented for scenario 3]{The top three most and least popular solutions for scenario three are listed along with their workshop ID's. In this scenario Susan cannot access bathrooms at the restaurant she delivers for. The (reverse) ranking averages across workshops are also offered to elucidate the strength of preference. Participants' votes on who should be responsible for changes is tallied. }
  \label{tab:s3}
\end{table*}

\subsubsection{Scenario 1 -- Lack of employment benefits (e.g. childcare, PTO)} \label{s1}

Worker and regulator participants preferred benefits such as childcare or part time off, which most workshops decided it was on platforms to implement. 
Platform-initiated development of childcare programs was considered especially ideal since it offers more flexibility in implementation, but the fear of receiving mandates does drive platforms towards action. In addition to childcare, paid time off can similarly offer temporary relief to Renee's situation. However, platforms were reluctant to provide benefits like these due to restricted funding. As non-employees, workers are currently not guaranteed allowances like paid time off or childcare support, and platforms fear that any government mandates requiring so might incur additional costs.
As an exception, regulators from Washington state have set an example for other localities by granting gig workers certain guarantees like sick leave or minimum wage, without sacrificing their status as an independent contractor\footnote{Bill HB2076 offers Washington drivers sick leave and minimum wage standards when they transport a passenger in their car: \url{https://lawfilesext.leg.wa.gov/biennium/2021-22/Pdf/Bills/House\%20Passed\%20Legislature/2076-S.PL.pdf?q=20220309063519}}. Finally, regulators and platforms were both inclined to avoid regulatory micromanagement, but welcome platform-initiated changes, which could be incentivized by regulations. One way to motivate rather than regulate platforms is through taxation mechanisms, where platforms either receive a tax break for providing a certain benefit, or pay the a tax for the government to provide the benefit to workers. Some platform designers may prefer this solution since a worker benefit program or service with regulation might mandate a specific timeline or a particular way of implementation.

\textbf{Summary of stakeholder stances and recommendations}: Everyone valued worker benefits (e.g., childcare and PTO) highly, and were inclined to think that platforms implement and pay for it. But platforms were reluctant to act due to associated costs and legal liabilities. Regulators can incentivize platforms by mandating some workers benefits, but should guard against micromanaging the execution of such initiatives.

\subsubsection{Scenario 2 -- Income instability} \label{s2}
Platforms are overwhelmingly happy to plan off-season events to help decommissioned workers, since it also brings them earnings. In fact, one participant's employer platform already offers an effective incentive program for workers to complete snow removal jobs. One way of encouraging client engagement that participants recommended was the initiation of a ``spring-cleaning week'', which would prompt them toward a task that they wouldn't otherwise think about. Such events advantage workers by giving them information that substitutes for the social network they would've relied on informally. However, workers worry that income from platform-initiated events offer only minor gains, not long-term solutions -- it was imperative to workers that they can plan for and control their own financial situations. One way that workers can curb the effects of seasonal fluctuations was to leverage the availability of multiple platforms, so that when they don't have work at TaskBunny, they can earn through jobs somewhere else. Platforms can also help workers conduct financial planning by including features like in-app earnings projections. Finally, platforms are disinclined to provide unemployment benefits, citing (on top of costs) how disbursing unemployment funds upfront may cause recipients to immediately spend it or lose their motivation to work.

\begin{table*}[h]
\caption{Scenario 4 Rankings and Voting Summary}
\begin{tabular}{|l|llc|}
\hline
\multirow{8}{*}{\begin{tabular}[c]{@{}c@{}}\includegraphics[width=2.75cm]{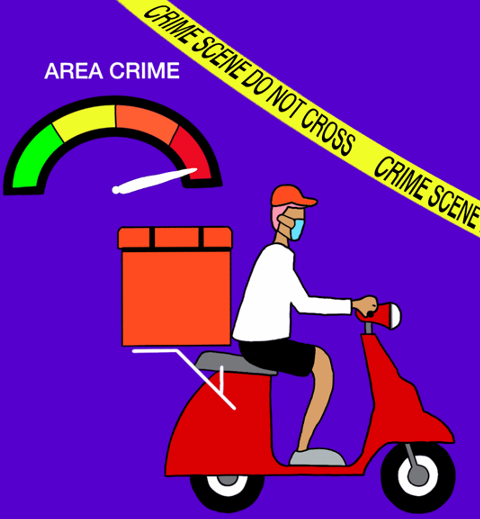} \\ \add{George receives a high} \\ \add{medical bill for injuries} \\ \add{received from an attack at} \\ \add{ an unsafe area after a delivery.}\end{tabular}} & \multicolumn{2}{c|}{\textbf{Scenario 4} \add{(Undermined Safety \& Worker Protections)}}                                & \textbf{Ranking} \\ \cline{2-4} 
& \multicolumn{1}{l|}{\multirow{3}{*}{\begin{tabular}[l]{@{}c@{}}Top 3 \\ most \textbf{favored}\\  solutions\end{tabular} }} & \multicolumn{1}{l|}{Regulators pass universal healthcare. {[}R2-3, W1-2, P2{]}} & 3.375 \\ \cline{3-4} 
& \multicolumn{1}{l|}{}                  & \multicolumn{1}{l|}{Platforms provide security equipment. {[}R1-2, W1, P3{]}} & 3.250 \\ \cline{3-4} 
& \multicolumn{1}{l|}{}                  & \multicolumn{1}{l|}{Platforms provide worker's compensation. {[}R1-2, W2, P2{]}} & 3.250 \\  \cline{2-4} 
& \multicolumn{1}{l|}{\multirow{3}{*}{\begin{tabular}[c]{@{}c@{}}Top 3 \\ most \textbf{disliked}\\   solutions\end{tabular}}} & \multicolumn{1}{l|}{\begin{tabular}[l]{@{}l@{}}
\begin{tabular}[l]{@{}l@{}}
Regulators restrict platforms from sending drivers to \\ high-crime areas.  {[}R2-3, W1-2, P1-3{]}
\end{tabular}
\end{tabular}} & 7.563 \\ \cline{3-4} 
& \multicolumn{1}{l|}{}                  & \multicolumn{1}{l|}{\begin{tabular}[l]{@{}l@{}}
Regulators require platforms to issue a warning \\ when workers enter high-crime areas. {[}R2-3, W1-2, P2{]} \end{tabular}} & 5.250 \\ \cline{3-4} 
& \multicolumn{1}{l|}{}                  & \multicolumn{1}{l|}{\begin{tabular}[l]{@{}l@{}}
Platforms provide workers additional subsidies  \\ for serving in high-crime areas. {[}R2-3{]}\end{tabular}} & 4.875 \\ \cline{2-4} 
& \multicolumn{1}{l|}{\begin{tabular}[c]{@{}c@{}}Who should be \\ responsible for \\ making changes\end{tabular} }                  & \multicolumn{2}{l|}{\begin{tabular}[c]{@{}l@{}}$\bullet$ 8 of 8 workshops voted platforms [R1, R2, R3, W1, W2, P1, P2, P3]\\ $\bullet$ 2 of 8 workshops voted workers [W1, P1]\\ $\bullet$ 6 of 8 workshops voted regulators [R1, R2, R3, W1, P2, P3]\end{tabular}}    \\ \hline
\end{tabular}
  \Description[A summary of stakeholders' responses to solutions presented for scenario 4]{The top three most and least popular solutions for scenario four are listed along with their workshop ID's. In this scenario George was attacked after a delivery and did not have health insurance to cover injuries. The (reverse) ranking averages across workshops are also offered to elucidate the strength of preference. Participants' votes on who should be responsible for changes is tallied. }
  \label{tab:s4}
\end{table*}


\textbf{Summary of stakeholder stances and recommendations}: Compared to platform-planned off-season events, workers preferred to be in control of their own financial planning. Since platforms were unwilling to provide unemployment benefits, workers can overcome seasonal lows by engaging with alternative platforms. Such worker inclinations toward increased agency presents unique opportunities for HCI designers to invent technological solutions for workers that integrate multiple platforms and facilitate cross-platform information sharing.

\subsubsection{Scenario 3 -- Missing Access to Working Necessities} \label{s3}
All workshops recognized bathroom access as a basic need. As service-providers to restaurants, workers (along with regulators) felt adamant that deliverers like Susan should not be denied necessary access to bathrooms. One worker was willing to publicly voice such opinions through petitions and suggested that platforms issue badges to workers so that they can be given direct bathroom access in restaurants.
While regulator participants conceded that public infrastructure improvements are needed to build more clean and safe bathrooms, they also believe it is platforms' responsibilities to negotiate with restaurants, and to share with the workers a map indicating restaurants where the public is allowed to use the restroom. Unfortunately, platforms were reluctant to require bathroom access for workers from restaurants because they predict a drop-off in the number of participating restaurants. One platform participant commented that it's really hard to make  bathroom access mandatory from the food safety perspective. On the other hand, a regulator also noted how there are health code requirements that expect bathrooms to be publicly accessible. Bathrooms are one instance of underdeveloped public service, and in general we find that gig work exposes a lack of basic, fundamental safety nets in our public infrastructure.

\textbf{Summary of stakeholder stances and recommendations}: Our existing public infrastructure does not offer enough safe and public bathrooms, and gig work is starting to probe at the social boundary between platforms, restaurants, and workers regarding how workers should access facilities like bathrooms that are essential for work. Platforms can offer technological support by integrating restroom locations into maps and incorporating restroom breaks into route planning.

\subsubsection{Scenario 4 -- Undermined Safety \& Worker Protections} \label{s4}
The idea of restricting deliveries in high crime areas was rejected by all three stakeholder groups. In particular, regulators discouraged investing in technological improvements (e.g. signals and buttons and alerts) because identifying dangerous locations can evolve into digital redlining, thereby reinforcing existing stigma surrounding the place. Cutting off orders hurts restaurants because it generates less revenue, harms drivers by reducing their income, and angers hungry people since they can't get food delivery. Regulators recognized how this scenario calls attention to underlying issues of unsafe communities, and to address these, all workshops voted for platforms to contribute toward community safety improvements, through provisions of a safe operational vehicle, personal protective equipment etc. But security measures shouldn't really mean just the equipment, it also involves security personnel, which can take the form of visible public presences such as the police. Unfortunately, the public police force in general is overstretched and underfunded. Even if emergency buttons directing to the police were to be implemented, they would be fraught with issues related to fair distribution -- people would wonder why higher status law enforcement is more responsive to the platforms and its drivers, raising questions like ``Why did GrubHub drivers get the button? Why doesn't everybody get a button?'' (R3). Worker and regulator participants also thought that platforms should provide workers' compensation, especially if the injuries were received in area where workers arrived to for a gig. From a worker's perspective, those compensations could go a long way in helping George pay for his bills. Lastly, a regulator suggested providing more available medical facilities so that workers can have ``a place where they can go and get that quick healthcare'' (R2).

\textbf{Summary of stakeholder stances and recommendations}: Segregating areas by restricting (delivery) services in  high-crime locations is not the way forward. Regulators and platforms should work together to improve community safety. In particular, platforms should invest in security equipment for workers while regulators can provide more visible public presences as security personnel. 

\begin{table*}[h]
\caption{Scenario 5 Rankings and Voting Summary}
\begin{tabular}{|l|llc|}
\hline
\multirow{8}{*}{\begin{tabular}[c]{@{}c@{}}\\ \includegraphics[width=2.75cm]{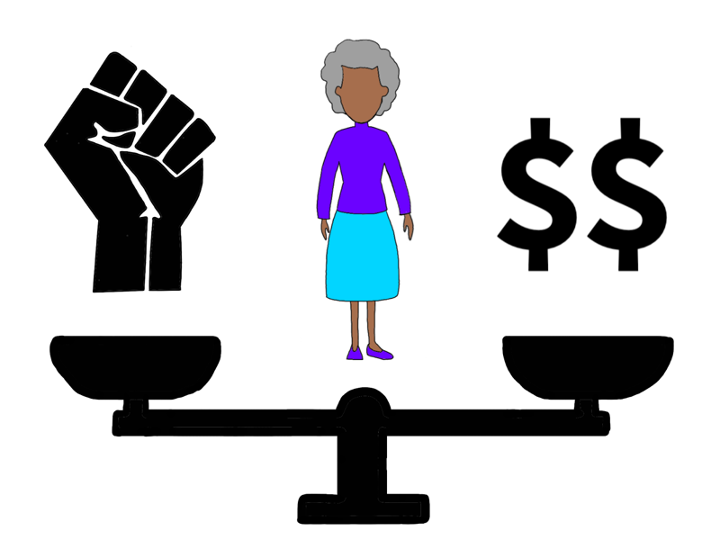} \\ \add{Marianne's earnings} \\ \add{were compromised} \\ \add{after intransparent and} \\ \add{unfair platform decisions}\end{tabular}} & \multicolumn{2}{c|}{\textbf{Scenario 5} \add{(Intransparency \& need for collective action)}}                                & \textbf{Ranking} \\ \cline{2-4} 
& \multicolumn{1}{l|}{\multirow{3}{*}{\begin{tabular}[l]{@{}c@{}}Top 3 \\ most \textbf{favored}\\  solutions\end{tabular} }} & \multicolumn{1}{l|}{\begin{tabular}[l]{@{}l@{}}
Platforms implement transparent policies \\ about decisions to keep workers informed. {[}W1, P1, P3{]}
\end{tabular}} & 3.750 \\ \cline{3-4} 
& \multicolumn{1}{l|}{}                  & \multicolumn{1}{l|}{\begin{tabular}[l]{@{}l@{}}
Workers notify buyers of their situation \\ to garner support. {[}R1-R3{]}
\end{tabular}} & 3.875 \\ \cline{3-4} 
& \multicolumn{1}{l|}{}                  & \multicolumn{1}{l|}{\begin{tabular}[l]{@{}l@{}}
Regulators impose a ceiling on transaction fees. \\ {[}R1, R3, W1{]}
\end{tabular}} & 4.250 \\  \cline{2-4} 
& \multicolumn{1}{l|}{\multirow{3}{*}{\begin{tabular}[c]{@{}c@{}}Top 3 \\ most \textbf{disliked}\\   solutions\end{tabular}}} & \multicolumn{1}{l|}{\begin{tabular}[l]{@{}l@{}}
\begin{tabular}[l]{@{}l@{}}
Workers pool savings to strike without losing income. \\ {[}R1-2, P1-2{]}
\end{tabular}
\end{tabular}} & 6.000 \\ \cline{3-4} 
& \multicolumn{1}{l|}{}                  & \multicolumn{1}{l|}{\begin{tabular}[l]{@{}l@{}}
Workers maintain a good relationship with platform \\ by not participating in the strike {[}R2-3, P2{]} \end{tabular}} & 5.625 \\ \cline{3-4} 
& \multicolumn{1}{l|}{}                  & \multicolumn{1}{l|}{\begin{tabular}[l]{@{}l@{}}
Workers participate in the strike by stopping sales. \\ {[}W1, P1-2{]}\end{tabular}} & 4.313 \\ \cline{2-4} 
& \multicolumn{1}{l|}{\begin{tabular}[c]{@{}c@{}}Who should be \\ responsible for \\ making changes\end{tabular} }                  & \multicolumn{2}{l|}{\begin{tabular}[c]{@{}l@{}}$\bullet$ 6 of 8 workshops voted platforms [R2, W1, W2, P1, P2, P3]\\ $\bullet$ 5 of 8 workshops voted workers [R1, R3, W1, W2, P2]\\ $\bullet$ 5 of 8 workshops voted regulators [R2, R3, W1, W2, P3]\end{tabular}}    \\ \hline
\end{tabular}
  \Description[A summary of stakeholders' responses to solutions presented for scenario 5]{The top three most and least popular solutions for scenario five are listed along with their workshop ID's. In this scenario Marianne's earnings were compromised as a result of a platform decision, which was not accurately reported to workers. The (reverse) ranking averages across workshops are also offered to elucidate the strength of preference. Participants' votes on who should be responsible for changes is tallied. }
  \label{tab:s5}
\end{table*}

\subsubsection{Scenario 5 -- Intransparency \& need for collective action} \label{s5}

Transparent policies were most desired by both worker and platform participants, so that sellers like Marianne have time to plan for drastic changes. Because Ebsy failed to communicate their decisions to workers like Marianne ahead of time, now she has to deal with the dilemma of whether or not to strike. Even platform employees thought Ebsy ``definitely did a wrong thing'' by destroying their ``long-term trust situation'' with sellers through intransparency, which is ``something we should avoid, and the regulators should require transparent policies \dots because sellers is actually why your platform exist[s]'' (P1). To help workers achieve financial stability, platform participants recommended sellers strengthen their portfolio by putting their products on different platforms. This strategy of multi-apping is commonly employed even before the pandemic, and across continents \cite{goods2019your}. 
Regulator participants heavily encouraged workers like Marianne to participate in collective actions such as strikes, citing a list of reasons: it is a way of gaining power, Marianne owes her coworkers the support, and because solidarity is what makes strikes work. 
However, regulators also acknowledged the difficulties of collective organization, since it requires a ``certain savvy with regard to using social media'' (R3), which requires careful planning as a community. Indeed, workers strongly resisted engaging in collective action (as is observable through the most disliked solutions), expressing that they did not feel ``comfortable having their savings pooled together'' (W2). One platform participant also recommended that workers refrain from striking and ``maintain a good relationship with Ebsy'' (P1), rationalizing that doing so can advantage Marianne by boosting her sales while other sellers strike. 

\textbf{Summary of stakeholder stances and recommendations}: Transparency is a good first step for ensuring that workers have agency in making alternative plans. However, collective actions can be complicated since gig workers are not legally categorized as employees, and hence cannot formally unionize. Furthermore, it's difficult for workers to build enough trust among one another to contribute toward pooling or strikes.

\subsection{Participant-generated Solutions}
In the following section we highlight some new ideas that participants organically generated during workshops. During the analysis phase, we divided these contributions into radical and reach solutions and further categorized them by the stakeholder group(s) that can bring them into reality. Table \ref{tab:new_solutions} summarizes these ideas while a full list of stakeholder-generated solutions can be found in the scenario-separated Tables \ref{tab:s1_sols} and \ref{tab:s4_sols} of the Appendix.

\begin{table*}[t]
\begin{tabular}{|l|lll|}
\hline
 &
  \multicolumn{1}{c|}{\textbf{Platforms}} &
  \multicolumn{1}{c|}{\textbf{Regulators}} &
  \multicolumn{1}{c|}{\textbf{Workers}} \\ \hline
 \multicolumn{1}{|l|}{\begin{tabular}[l]{@{}l@{}}
\textbf{Radical \slash} \\ \textbf{Reach} \\ \textbf{Solutions} \end{tabular}}
&
  \multicolumn{1}{l|}{\begin{tabular}[c]{@{}l@{}}  
  $\bullet$ Partnerships between platforms \\ $\bullet$ Improved transparency policies \\ $\bullet$ Cross-platform worker rating system\\ $\bullet$ Green light hubs \\  $\bullet$ Platform-subsidized maternity leave \end{tabular}} &
  \multicolumn{1}{l|}{\begin{tabular}[c]{@{}l@{}}$\bullet$ A third legal class of workers\\ $\bullet$ More clean \& safe public \\ bathrooms \\ $\bullet$ More police \slash safety solutions\end{tabular}} &
  \multirow{2}{*}{\begin{tabular}[c]{@{}l@{}}$\bullet$ Worker-owned \\cooperatives\\ $\bullet$ Worker-initiated \\ petitions \& strikes\end{tabular}} \\ \cline{2-3}
 &
  \multicolumn{2}{l|}{\begin{tabular}[c]{@{}l@{}}$\bullet$ Mandatory company-funded worker compensations\\$\bullet$ Regulator/platform-backed income pools \\$\bullet$ Universal basic income\\ $\bullet$ Higher hourly pay for all\\ $\bullet$ Improved insurance schemes \\ $\bullet$ Price ceiling on all transactions \\ \end{tabular}} &
   \\ \cline{2-4} 
 &
  \multicolumn{3}{l|}{$\bullet$ Shifts in legal and social classifications of gig workers} \\ \hline
\multicolumn{1}{|l|}{\begin{tabular}[l]{@{}l@{}}
\textbf{Incremental} \\ \textbf{Changes} \end{tabular}}
&
  \multicolumn{1}{l|}{\begin{tabular}[c]{@{}l@{}}  $\bullet$ Reduce wait times \& offer better rides \\ $\bullet$ Allow worker-scheduled rides\\$\bullet$ Earnings projections with category \\ suggestions\\ $\bullet$ Company-supported savings\\ $\bullet$ Employer-sponsored financial education\\ $\bullet$ Worker-success programs\\ $\bullet$ Trust-based loans \& loyal worker bonuses\\$\bullet$ Within-vehicle lock mechanisms\\ $\bullet$ Emergency button on bikes\\  $\bullet$ Anti-violence investment\end{tabular}} &
  \multicolumn{1}{l|}{\begin{tabular}[c]{@{}l@{}} $\bullet$ Employee assistance \\
  programs (EAPs)\\ $\bullet$ Job training\\ 
  {\begin{tabular}[l]{@{}l@{}}
    $\bullet$ Help workers connect with \\ the local workforce system \end{tabular}}
  \end{tabular}}    &
  \begin{tabular}[c]{@{}l@{}}$\bullet$Leverage multiple \\ platforms\\ $\bullet$ Make financial \\ plans personally \end{tabular} \\ \hline
\end{tabular}
\caption{Participant generated solutions}
\label{tab:new_solutions}
\Description[Outline of raw, participant-generated solutions, where one axis delineates the different stakeholder groups whereas the other categorizes whether they are incremental or radical efforts.]{The vertical axis of the table divides the incremental solutions from more radical or reach solutions. The horizontal axis labels which stakeholder groups generated the solutions. Many of the reach solutions were generated by multiple stakeholder groups, these were indicated by merged cells.}
\vspace{7mm}
\end{table*}

\subsubsection{Radical Re-imaginings}

\paragraph{Platform-side Actions} \label{platform_radical}
Many platform stakeholders consider \textbf{multi-platform partnerships} plausible and effective solutions. Discounts for childcare was one partnership idea from P2, which would work ``if there was some childcare provider, and [with them as a partner] we said [to workers] because you're a worker [on our platform], you get 60\% off or something'' (P2). Help with rent is another benefit that partnerships can provide workers, where they receive ``a \$20 contribution that could be used then on this GigEasy platform to purchase rent protection'' (P2). Finally, P3 imagined a cross-platform rating system for workers so that their reputations can be shared across platforms, which can allow workers to easily maintain reputation across platforms and for platforms to recommend workers to one another.

All stakeholder groups advocated for \textbf{improved platformic transparency}, which can help increase worker autonomy and agency. For instance, one platform designer conjectures that ``if you presented it [earnings projections] in the right way and maybe said: `you're tasking in the moving category and we expect like during these months, this will be your earnings. But here's some categories where we think this would be your earnings and you should sign up for those' ''(P2), then workers would have more options on improving income. \textbf{Well-presented, transparent, and actionable recommendations} would offer workers insights for long-term planning. 

On top of technological improvements, \textbf{platforms can help alleviate the shortcomings of public infrastructure}. For instance P2 called for the establishment of more green light hubs, or partner support centers that contain lounges and bathrooms, so that workers can have physical locations to stop, rest and support one another. W1 and R2 both organically generated universal maternity leave (paid for by companies) as a solution for Renee, and W1 even voted for it as their favorite solution. 

\paragraph{Regulatory Actions} \label{revisions}
Taking a more revisionist approach, a P2 participant envisioned ``\textbf{a third legal class of worker}[s] existing'', since ``so much of the legal battle has been about: either you're a contractor or you're an employee \dots if there were some third class of worker, then you could actually have an employment scheme that made sense for the type of work that people were doing''. By shifting the focus of platforms away from the legal risks of overstepping the boundaries of contractual work, a new classification could redirect efforts toward more improvements and protections.

The previously unprecedented rise in gig work revealed \textbf{numerous inadequacies in our public infrastructure}, where many fundamental improvements are needed to ensure the sustainable functioning of the gig economy. Both R3 participants vehemently stood up for ``more clean, safe public bathrooms'' and P3 thought the government should send more police (or safety solutions) to help unsafe neighborhoods for cases like George's.

\paragraph{Worker-side Actions}
Many regulators supported worker-initiated \textbf{petitions, strikes (\ref{worker_action}) and  worker-owned cooperatives} (R2). But while collective efforts are easier to introduce than new regulations ``because it doesn't require any sort of legal intervention'', collective organization is difficult since ``most of the people I know who drive \dots they don't want that kind of responsibility'' (R3). Platform themselves act as an additional barrier against community-building, since they ``intentionally never \dots built up any type of community around the drivers'' (P2). 

\paragraph{Collaborations Between All Stakeholders} \label{all}
Instead, participants proposed \textbf{shifts in legal and social classifications of workers} \cite{muntaner2018digital} since ``gig worker[s] these days \dots are treated in a variety of political ways, legal ways, social ways, cultural ways \dots and so we, as a matter of public policy \dots should be figuring out how to level it up'' (R3). Improved treatment of gig workers can start from us all, by ``changing our preconception about who a worker is, and what it means to work, and the kind of vulnerabilities that you have as a worker in a gig economy'' (R3), we would collectively contribute toward improved perceptions of and conditions for gig workers. 

\paragraph{Co-regulated Platformic/Government Actions} \label{coregulate}
While a legal reclassification of workers can help them reap many benefits and protections, such drastic labor law adjustments are unlikely to take effect in the near future. In the meantime, regulators and platform designers recommended more specific \textbf{policies to protect worker safety and earnings}. For cases like George, R1 advocated for mandatory company-funded worker compensations (to ameliorate the costs of task-related injuries), and R3 suggested regulator/platform-backed income pools for seasonal workers like Dave.

Beyond policy revisions and additional mandates, participants also advocated for more radical and reach solutions that provide \textbf{universal benefits}, while acknowledging their current infeasibility. For instance, universal healthcare (a researcher-generated idea), garnered the most support and was the highest ranked solution across three workshops for George's scenario. R2 participants proposed earning guarantees such as universal basic income for Dave's situation, and higher hourly pay for everyone in the case of Renee. P2 recommended improved insurance schemes with a fixed coverage gap and R1 advocated for the government to impose a price ceiling on all transactions to reduce the risks that sellers like Marianne experience wage theft.

\subsubsection{Incremental Improvements}

\paragraph{Platformic Actions}
To build upon existing algorithmic functions, participants proposed various \textbf{new platform features and initiatives to help workers improve efficiency, raise earnings and protect health and safety}.
To approach higher worker productivity, P3 recommended optimizing the existing algorithm to reduce wait times, offer better rides/tasks, and allow workers to schedule rides ahead of time. 
To increase earnings, participants suggested new category suggestions (P2) and company supported savings (R2).
More indirectly, workers can raise their earnings by acquiring or honing (new) skills. Hence, participants recommended initiatives such as employer-sponsored financial education (R3) and worker-success programs (P2) so that workers can adjust for marketing offerings, availabilities, supplies, etc. For veteran workers, trust-based loans or bonuses (P2) can dissuade loyal workers from leaving the platform. 

Participants generated a variety of ways that platforms can help promote physical safety. Some ``quick hit, easy solution[s]'' include a ``\textbf{locking mechanism in the vehicle} \dots a drop space you can't open, [because] more than once, I've known day workers getting mugged because they're easily identifiable as having money on them'' (R2) as well as  ``\textbf{driver check-ins and an emergency button} \dots it's not gonna get [to] the root cause, \dots [but it is a] small way to assure that the workers feel a little bit more comfortable'' (R3). A W1 worker also confirmed  prior findings of driver preferences on safety equipment \cite{Almoqbel2019-in}, stating that ``driver check-in also is good  \dots just in case things like attacks happened''. Finally, platforms can begin ``\textbf{investing in that kind of root cause anti-violence work} that the particular municipality or locality might need \dots [which] could be [delivered] in the form of a grant to that municipality'' (R3). 

\paragraph{Regulatory/Government Action}
Many of these aforementioned programs and benefits are also implementable by governments. For instance one R1 participant pointed out how \textbf{employee assistance and job training programs} already exist. Meanwhile, \textbf{helping workers connect with local workforce system} could have assisted workers like Dave seek additional tasks and income during off seasons.

\paragraph{Worker-side Actions}
In addition to changes from the platform end, participants also suggested ways that workers can take the matter into their own hands. W1, W2 and P3 all recommended workers like Marianne to \textbf{leverage multiple platforms} by selling products on these different sites simultaneously (\ref{s5}), so as to curb the effects of unforeseen situations. In the case of Dave, W2 participants saw an opportunity for the worker to make financial plans personally to prepare for the effects of seasonal changes.

\section{Discussion}
In this study, we took a stakeholder-driven approach with platforms, regulators and workers to examine pressing issues related to gig work. 
In doing so, we hope to provide a richer and more holistic picture of where we currently stand in terms of gig work conditions, as well as where improvements are possible and most needed.
By conducting these co-design workshops with relevant stakeholder groups, we can address a broader set of needs, approach more practical and realistic designs, and further our progress in creating the gig work futures that we discuss, imagine, and dream for together. 
In the following section, we shed light on these multi-stakeholder findings by highlighting design recommendations, ideas for collaboration, and key insights that emerge from the intersection of stakeholders' perspectives. 
On top of recommending new avenues for future work and developments in service, policy and technology, we also provide cautions against potentially harmful side effects that may arise from implementing these solutions.
\subsection{Implications for Technological Developments}
\begin{itemize}
  \item \textbf{Platform-initiated changes as low hanging fruits}. Our findings suggest that platforms can initiate a plethora of incremental changes for improving gig work conditions, including ways of increasing earning/financial opportunities as well as additions for benefiting worker health and safety. For instance, the in-app display of public bathroom locations was one of the most favored solutions in \ref{s3}, and may serve as a temporary fix for the current shortage of public bathrooms. To help curb the seasonal nature of gigs, platforms can recommend off-season work opportunities and provide in-app earnings projections to guide financial planning (\ref{s2}). Such features are also aligned with platforms' overall preferences and can benefit platforms in the long run, by offering competitive advantages that help to retain existing workers and attract newcomers.   
  \item \textbf{Technologies that motivate workers to voice concerns without harming earning opportunities}. Currently, workers hesitate to engage in collective actions despite overwhelming support from advocates and regulators because they 1) lack legal protections and social support and 2) fear a loss of work opportunities that may result from damaged relationship with platforms. Future system designers can explore ways of encouraging prosocial data-sharing among workers to foster communities of support, where workers can protect and advocate for their gig community's well-being with data-driven insights without needing to worry about legal implications or reputational consequences \cite{calacci2022organizing}. Prior studies have suggested using data-driven insights to raise public awareness about worrying circumstances surrounding (gig) work environments \cite{Khovanskaya2019-xo, calacci2022organizing}, and a feasibility analysis showed the potentials of platform cooperatives replacing investor-owned platforms \cite{bunders2022feasibility}. 
  Mobilization of gig workers have also occurred at increasing rates in Europe \cite{cini2022or} and Latin America, where they leveraged social media to coalesce in large-scale, organized, international strikes \cite{howson2020just}, proving how informal labor networks and mutual aid can transform distributed workforces even in the absence of formal union structures \cite{qadri2021s}. 
  \item \textbf{Multi-platform collaborations}. Gig platforms largely coexist as competitors to one another. Our participants encouraged multi-platform collaborations, which can benefit both workers and platforms. For example, partnerships across platforms can help workers battle the instabilities of gigs (\ref{s2}) and provide assistance with childcare (\ref{s1}) while cross-platform worker ratings can encourage to workers reuse a single portfolio across platforms and tasks (\ref{platform_radical}), which can increase earning opportunities (\ref{multiapp}, \ref{s5}) \cite{hsieh2022little}. 
  Recent work anticipates the need for both workers and clients to engage the services of several platforms simultaneously, pointing to potential rise of multi-platform systems \cite{amiri2021separ}. This suggests an opportunity gap where tooling and resources can be developed to help workers easily transition and switch between platforms.
\end{itemize}

\paragraph{Cautions}
The innovations proposed above can have potentially deleterious side effects that developers should guard against. For instance, a system for collective actions can \textbf{expose and breach the privacy of protesting workers}, possibly causing losses of earning opportunities. Furthermore, while our workers called for more personalized accommodations, such arrangements inevitably \textbf{trades off with privacy }\cite{sannon2022privacy, garcia2016personalization, lee2013designing}, potentially requiring platforms to access and monitor working habits and other behaviors. Implementations of personalization features should take care to not cross the line between customization and invasive surveillance. Finally, the cross-platform ratings of workers can exert \textbf{overt pressure on workers to maintain good reputation} -- small disagreements with one client could affect their earning potentials across multiple platforms. Hence, designers of multi-platform rating systems should consider protective mechanisms that prevent clients from abusing their rating privileges. 

\subsection{Implications for Policy Advancement}
\begin{itemize}
    \item \textbf{Regulations to incentivize platform-initiated programs and accommodations.} While regulators strongly advocate for empowering the collective voice of gig workers and creating better gig work environments, platforms are reluctant to provide such resources, listing a plethora of reasons for such inaction.  Hence, policymakers and platforms should work together to devise regulatory measures that motivate platforms to mobilize and provide services/resources that benefit worker well-being. Such incentives can take many forms: our participants suggested tax breaks (\ref{s1}), government subsidies (\ref{platform_radical}), and in the case of Washington state -- new litigation to offer benefits such as workers' compensation alongside the flexibility of independent contracting (\ref{s1}) \cite{noauthor_2022-wo}. 
    \item \textbf{Regulations on platforms to ensure  occupational health \& safety.} Many of our regulator participants admit that some of the occupational risks enburdening gig workers in the US are consequences of missing or inadequate public infrastructure. For example, the lack of available public bathrooms contributed to Susan's inability to meet a dire biological need at work (\ref{s3}), and this shortage has only been aggravated by the pandemic \cite{bathroom}. Similarly, physical safety of food couriers can be compromised in the wake of rising crime without protections by visible public presences (\ref{s4}) \cite{assault}. Thus, it is of increasing urgency for policymakers to propose mandates and regulations to drive platforms' efforts that promote gig worker health and safety and subsequently for regulators enforce such directives, so as to close the gap between policy and regulation \cite{fan2022online}.
    \item \textbf{Enhanced legal \& public perceptions of gig work.} As Howard found, the legal misclassification of gig workers as contractors is a major contributor to their substandard conditions of occupational health and safety \cite{Howard2017-wd}. Participants brought up both legislative and cultural shifts in how we consider gig workers (see \ref{revisions} \nameref{revisions} and \nameref{all}) as first steps toward mitigating existing social stigmas and legal misclassifications. That is, a change in worker status must begin with an updated perception of workers from the public at large -- we should raise our own awareness of workers' vulnerabilities instead of considering them as fungible/replaceable, and reflect on how we can contribute toward improvements of current conditions. While an abundance of reports and studies have criticized how platforms abuse the inappropriate classification of gig workers as contractors to subvert corporate responsibilities and liabilities \cite{lobel2017gig, Gray2019-ue, Dubal2017-bj, de2015rise, tran2017gig, delfino2018work}, further advancements in policy and public discourse are needed to provide workers with the employee benefits and protections they deserve.
\end{itemize}

\paragraph{Cautions}
An excess of specific regulations run the risk of \textbf{micromanaging platforms} (\ref{s1}), therefore regulators should provide companies enough flexibility in how they implement benefit programs and services to workers, but at the same time make sure the changes are measurable and enforceable, as Johnston et. al. suggested \cite{Johnston_undated-wy}. Regarding proposed improvements for public infrastructure (e.g. bathroom access and public safety), regulator participants expressed concerns around \textbf{redlining districts that are less safe or developed}, hence future policy proposals should be inclusive of traditionally underserved populations and localities \cite{tran2017gig, Dillahunt2018-ia, graham2018could}.

\subsection{Implications for Service and Management Practices}
\begin{itemize}
    \item \textbf{Regulators and platforms prioritize \& co-regulate (universal) benefits.} Regulator and worker participants welcomed various forms of employee benefits, including healthcare, security equipment, worker's compensation, price ceilings on transaction fees, and childcare services (Table \ref{tab:s3} and \ref{tab:s1}). Many of these ``universal'' benefits require the co-regulation from regulators, lawmakers and platforms, so that they can collaborate in fixing legal loopholes and market inefficiencies (\ref{coregulate}) \cite{cannon2014framework}. Hence, future work can investigate ways of measuring the costs and returns of implementing the various types of employee benefits, so that legal and platform practitioners can better prioritize services to meet worker needs.
    \item \textbf{Green light hubs / worker rest areas. }The temporary nature of gigs makes workers lack many forms of physical support, and inadequacies in our public infrastructure lengths their already extensive list of occupational hazards \cite{tran2017gig}. While we can hope that gig work speeds up the development of these public sector services, there are no such guarantees in the near future. As an alternative, participants suggested for platforms to build more green light hubs \footnote{\url{https://www.ridester.com/uber-greenlight-hub/}} to provide workers physical locations for rest and (mutual) support (\ref{platform_radical}).
    \item \textbf{Follow worker recommendations in redesigns.} Conversations with diverse stakeholder groups increase our chances of addressing a broader set of needs and enables us to approach more practical and realistic designs, since redesigns of interactions between platforms and workers should involve \textbf{conversations between platforms and workers}. One worker pointed out how ``Renee interacts everyday with Lyber, and so the solutions need to come from their interactions'' (W1). As future platform designers and legislators work towards meeting the needs of workers, they should take heed to directly involve gig workers voices in the redesign process.

\end{itemize}


\subsubsection{Cautions}
In ranking and prioritizing worker benefits and programs, platforms and regulators may default to \textbf{short-term and unreliable solutions} as low-hanging fruits, which workers rejected. Hence, designers and providers should focus on the development of sustainable and reliable benefits/service offerings. 
On the other hand, there is a risk of \textbf{further encumbering workers with additional labor of devising solutions} for their own problems. Instead, collaborators should prepare optional solutions for gig workers to choose from when involving them in redesigns.

\subsection{Limitations, Reflections \& Future Work} \label{limits}

\subsubsection{Recruiting \& Participants}
Firstly, due to constraints of time and recruiting, we did not conduct these workshops with an extensive quantity of participants. Thus, we look forward to further collaborations with stakeholders from more diverse backgrounds. Second, we recognize that our method of recruiting of workers and platform designers online from Reddit or LinkedIn may exclude those who don't engage as regularly with online forums, such as the elderly population. In the future, we encourage designs to explore alternative recruitment techniques to mitigate this sampling bias.
Third, we refrained from involving multiple stakeholders in the same workshop due to potential power asymmetries among participants, (see \ref{recruit}), but acknowledge that this decision limits the results of our study by preventing discussion of differences in opinion across stakeholder groups. We invite future study designs to explore ways of avoiding such power differentials in a way that allows for the exchange of opinions among multiple stakeholders. 
Finally, our workshops were set in the context of the US, and while many of our platform-specific findings generalize at the global scale, policies implications may vary. We encourage future studies to investigate policy changes needed across nations, as well as possibilities for international regulations.

\subsubsection{Reflections}
While this methodology of interviewing multiple stakeholder groups offers an enriching and holistic perspective on issues related gig work conditions, we do acknowledge the broad nature of the presented issues. Follow up investigations could focus on particular issues or the evaluations of specific design ideas so as to elicit more actionable feedback from stakeholders. In particular, future efforts may continue exploring the possibilities of building collective bargaining power through data-sharing \cite{carpenter2023cagecoach, zhang2023stakeholder, calacci2022organizing, stein2022workers, calacci2022bargaining}.

The storyboards helped participants gauge the desirability of imagined futures, and afforded us the unique opportunity to ``rapidly investigate many possible futures'' by supporting ``a broad investigation of contexts, triggers, and interactions.'' \cite{Zimmerman2017-rq, Yoo2010-mu}, and we intentionally chose to depict scenarios that focused on individuals' challenges to encourage our participants to empathize with workers' struggles in each story. While many of the participants expressed a fondness for the scenarios, this decision may bias focus toward individual solutions. We encourage future work to frame scenarios and solutions as structural issues so as to generate more systematic solutions. Furthermore, we noticed that the average rankings were difficult to compare across scenarios since the varying number of pre-generated solutions caused the scale to differ. If future researchers choose to adopt the method, we encourage a uniform number of pre-populated solutions across scenarios.

Participants also commented on how considerations of these issues have benefited their day-to-day work. Notably, the councilperson from R3 mentioned how scenario 3 reminded them of something to bring up in an upcoming board meeting. For workers, we hope that the strategies we generated together can help face and overcome the issues we discussed in their daily work. Many participants also pointed out how some scenarios may not be representative of issues that the majority of workers face. We did intentionally present provocative problems so as to follow the speed dating method, but future study designs might consider involving more workers in the problem-generation stage to capture a problem space that more accurately represents the typical worker's day-to-day.

One of the desiderata we uncovered from workers was the enactment of more personalized solutions and policies of support. However, this preference conflicts with policymakers' objectives of treating all populations equally. One next step is to examine how existing labor laws may fall short in equitably serving the needs of gig workers, such explorations could consider how new policies and litigation can be implemented to incentivize platforms to provide workers the support and services they need. One effort in Europe has already started inquiring into workers data access rights \cite{stein2022workers}, and we encourage future researchers from across the globe to explore the impacts of local labor laws, so as to create more productive gig work ecosystems that empower workers to thrive in this new world of work. 

\section{Conclusions}
We took a first step to delve into the design of alternative futures for the gig workforce in a way that holds policymakers and platforms accountable for promoting fairer and healthier systems. From our workshops, we uncovered rationales behind why platforms resisted implementing worker benefits and protections, discovered regulators' strong support for helping gig workers build a collective voice, as well as workers' preferences for more individualized solutions. 
Our findings have practical implications for the future of worker advocacy, management practices as well as technological interventions, and we hope that future lines of research can extend these efforts to further advance the career trajectories of workers within the gig economy.

\begin{acks}
This material is based upon work supported by the National Science Foundation (NSF), in part by Award 1952085,
and in part by Grant No. DGE2140739. We are especially thankful to Karen Lightman from Metro21: Smart Cities Institute, for generously connecting us to stakeholder participants from the regulator group. We appreciate early advice provided by Min Kyung Lee, Angie Zhang, Zhi-Li Zhang, as well as Yanhua Li, which guided the construction of the storyboards. We would also like to thank audience feedback from CMU S3D's SSSG seminar as well as the UXA Flavors of HCI series. Finally, we are grateful to Seyun Kim, Franky Spektor, Motahhare Eslami, and anonymous reviewers for their invaluable feedback on various stages of the manuscript. Finally, we are grateful to our participants for their time and thoughtful insights during the workshops, as well as trust in the directions of our work.
\end{acks}

\bibliographystyle{ACM-Reference-Format}
\bibliography{references.bib}
\appendix
\begin{table*}[!t]
\section*{Appendix}
\caption{Participant-generated Solutions for Scenarios 1-3}
\Description[A table listing participant-generated solutions for the first three scenarios]{Within each scenario, the rows separate the participant-generated solutions by stakeholder group. }
\label{tab:s1_sols}
\begin{tabular}{|ll|}
\hline
  \multicolumn{2}{|c|}{\textbf{Scenario 1}} \\ \hline
  \multicolumn{1}{|l|}{Workers} &
  Platforms add policies for maternity leave, etc \\ \hline
\multicolumn{1}{|l|}{Platform employees} &
  \begin{tabular}[c]{@{}l@{}}Partner childcare program at discount\\ Third class of workers (not employees or contractors)\\ Bonuses for loyal workers (e.g. after 20 tasks)\\ Trust-based loans\\ Schedule rides in advance* \\ Platform provides better rides\end{tabular} \\ \hline
\multicolumn{1}{|l|}{Regulators} &
  \begin{tabular}[c]{@{}l@{}}Higher hourly pay to all\\ On-site health clinic\\ Universal maternity leave\\ Employee assistance program\\ Platforms consult driver focus groups (co-op)\\ Unionized labor\\ Flexible scheduling*\\ Collective bargaining\\ Nontraditional 24-hr child-care\end{tabular} \\ \hline
  \multicolumn{2}{|c|}{\textbf{Scenario 2}} \\ \hline
\multicolumn{1}{|l|}{Workers} &
  \begin{tabular}[c]{@{}l@{}}Find other gig work\\ Workers make personal plans for seasonal changes\end{tabular} \\ \hline
\multicolumn{1}{|l|}{Platform employees} &
  \begin{tabular}[c]{@{}l@{}}Marketing campaign\\ Train for other categories*\\ Projections with new category suggestions\\ 3rd class of workers\\ External tools like GigEasy for rent protection\\ Cross-platform gig worker ratings to help workers\end{tabular} \\ \hline
\multicolumn{1}{|l|}{Regulators} &
  \begin{tabular}[c]{@{}l@{}}Unions\\ Employee assistance program\\ Job training for other categories*\\ Connect with local workforce system\\ Inter-app collaborations \\ Universal basic income\\ Higher pay\\ Guaranteed minimum\\ Company-supported saving\\ Income pool (regulator/platform backed)\\ Employer-sponsored financial education\end{tabular} \\ \hline
  \multicolumn{2}{|c|}{\textbf{Scenario 3}} \\ \hline
\multicolumn{1}{|l|}{Workers} &
  Platforms offer/suggest breaks \\ \hline
\multicolumn{1}{|l|}{Platform employees} &
  \begin{tabular}[c]{@{}l@{}}Regulate more clean, safe public bathrooms*\\ Partnering with anyone (including clients) for public bathroom access\\ More green light hubs\\ Improve algorithms to reduce wait times\end{tabular} \\ \hline
\multicolumn{1}{|l|}{Regulators} &
  \begin{tabular}[c]{@{}l@{}}Cities add more bathrooms*\\ Regulations/unions enforce/bargain for less monitoring/oversight\end{tabular} \\ \hline
  \end{tabular}
\end{table*}

\begin{table*}[]
\caption{Participant-generated Solutions for Scenarios 4-5}
\Description[A table listing participant-generated solutions for scenarios 4 and 5]{Within each scenario, the rows separate the participant-generated solutions by stakeholder group. }
\label{tab:s4_sols}
\begin{tabular}{|ll|}
\hline
\multicolumn{2}{|c|}{\textbf{Scenario 4}} \\ \hline
\multicolumn{1}{|l|}{Workers} &
  \begin{tabular}[c]{@{}l@{}}More security personnel*\\ Bike emergency button\\ Ban bikes in high crime areas\\ Emergency button at all times\end{tabular} \\ \hline
\multicolumn{1}{|l|}{Platform employees} &
  \begin{tabular}[c]{@{}l@{}}Platforms notify workers when enter high crime areas\\ Improved insurance schemes with fixed coverage gap\\ Gov sends more police personnel to unsafe areas*\end{tabular} \\ \hline
\multicolumn{1}{|l|}{Regulators} &
  \begin{tabular}[c]{@{}l@{}}Health support\\ PTO\\ Worker compensation mandatory\\ Within vehicle lock mechanism (e.g. safes)\\ Platforms investing in anti-violence work\\ Social workers in place of cops\\ Visible public, physical presences\end{tabular} \\ \hline
  \multicolumn{2}{|c|}{\textbf{Scenario 5}} \\ \hline
\multicolumn{1}{|l|}{Workers} &
  \begin{tabular}[c]{@{}l@{}}Use another platform*\\ Contracts requiring certain fixed priced commodities (despite changes)\end{tabular} \\ \hline
\multicolumn{1}{|l|}{Platform employees} &
  \begin{tabular}[c]{@{}l@{}}Platforms implement better transparency policies\\ Worker-success program (adjusting marketing offerings, availabilities, supplies, etc)\\ Better supply-side quality control\\ Marianne puts her products on multiple platforms*\end{tabular} \\ \hline
\multicolumn{1}{|l|}{Regulators} &
  \begin{tabular}[c]{@{}l@{}}Connect to entrepreneur resources\\ Worker owned co-op\\ Support loyal workers via increased visibility\\ Marianne sells elsewhere*\\ Price ceiling on all transactions\end{tabular} \\ \hline
  \end{tabular}
\end{table*}

\normalsize
\begin{table*}[h]
\caption{(Researcher-generated) Solution Space}
\Description[A table listing the researcher-generated solutions.]{Solutions are first broken down by scenario, and then further by the corresponding stakeholder group that is responsible for implementation.}
\label{tab:solutions}
\begin{tabular}{|p{1cm}|lll|p{1.5cm}|}
\hline
\textbf{Scenario No.} &
  \multicolumn{3}{c|}{\textbf{Solutions}} &
  \multicolumn{1}{p{1.5cm}|}{\textbf{Responsible Group}} \\ \hline
\multirow{3}{*}{\textbf{1}} &
  \multicolumn{3}{l|}{\begin{tabular}[c]{p{11.5cm}}Lyber offers childcare programs/services to single parent drivers.\\ Lyber tells Renee the destination of her incoming rides so she can check in on her child when nearby.\\ Lyber provides guaranteed work during the day to single parents like Renee.\\ Lyber offers higher hourly pay to single parent drivers.\end{tabular}} &
  Platform \\ \cline{2-5} 
 &
  \multicolumn{3}{l|}{\begin{tabular}[c]{p{11.5cm}}Renee adds incentives like a single-mom badge to her profile/car to encourage more tips.\\ Renee joins a single mom driver group, and they will help each other when their kids are sick.\end{tabular}} &
  Worker \\ \cline{2-5} 
 &
  \multicolumn{3}{p{8cm}|}{Regulators require Lyber to provide days of PTO for all regular drivers, plus additional days for those with special needs such as single moms.} &
  Regulator \\ \hline
\multirow{3}{*}{\textbf{2}} &
  \multicolumn{3}{l|}{\begin{tabular}[c]{p{11.5cm}}TaskBunny recommends winter side-hustles (snow plowing/shoveling, food delivery, etc).\\ TaskBunny offers employer-sponsored contributions to help workers save and plan for long-term spending, even retirement planning.\\ TaskBunny projects earnings in-app to help long-term financial planning.\\ TaskBunny plans events during off-seasons to offer tasks for workers.\end{tabular}} &
  Platform \\ \cline{2-5} 
 &
  \multicolumn{3}{p{11.5cm}|}{Dave starts long-term financial planning using apps to budget, track spending, and plan for retirement.} &
  Worker \\ \cline{2-5} 
 &
  \multicolumn{3}{p{11.5cm}|}{Regulators provide unemployment benefits to contractual workers when they haven't received short-term jobs for XX months.} &
  Regulator \\ \hline
\multirow{3}{*}{\textbf{3}} &
  \multicolumn{3}{l|}{\begin{tabular}[c]{p{11.5cm}}Platforms should negotiate with restaurants to open bathroom locations to workers. Locations would be made available on the app (via built-in maps). \\ Platforms will show locations with public bathroom access.\\ Restaurants reduce wait times by cutting off online orders at popular hours.\\ Susan can request time for paid bathroom breaks from the app (at limited quantities per day).\end{tabular}} &
  Platform \\ \cline{2-5} 
 &
  \multicolumn{3}{l|}{\begin{tabular}[c]{p{11.5cm}}Susan shares her list of bathroom locations with other workers.\\ Workers band together to petition restaurants to give bathroom access (badges or QR code) to deliverers who serve their customers.\\ Workers report bathroom service quality (in-app or on another platform)\end{tabular}} &
  Worker \\ \cline{2-5} 
 &
  \multicolumn{3}{l|}{Regulators can require restaurants to provide bathroom access.} &
  Regulator \\ \hline
\multirow{3}{*}{\textbf{4}} &
  \multicolumn{3}{l|}{\begin{tabular}[c]{p{11.5cm}}Lyber offers George worker compensation even though he was offline during the time of attack.\\ Lyber gives workers additional subsidies for serving in high-crime areas.\\ Lyber provides security equipment (cameras, dash cams, etc) for drivers to turn on in high-crime areas while warning riders that they may be monitored. \\ Lyber checks in on drivers in between orders\end{tabular}} &
  Platform \\ \cline{2-5} 
 &
  \multicolumn{3}{p{11.5cm}|}{George should use Lyber's in-app emergency button during encounters of violence so that the police and ambulance can be dispatched.} &
  Worker \\ \cline{2-5} 
 &
  \multicolumn{3}{l|}{\begin{tabular}[c]{p{11.5cm}}Regulators require platforms to issue a warning when workers enter high-crime areas. \\ Regulators should restrict platforms to send drivers to high-crime areas.\\ Regulators pass universal healthcare\end{tabular}} &
  Regulator \\ \hline
\multirow{3}{*}{\textbf{5}} &
  \multicolumn{3}{l|}{\begin{tabular}[c]{p{11.5cm}}Ebsy should issue an apology and compensate long-time sellers for being intransparent about the directions of their funding\\ Ebsy implements new transparency policies about their decision making so workers can be better informed\end{tabular}} &
  Platform \\ \cline{2-5} 
 &
  \multicolumn{3}{l|}{\begin{tabular}[c]{p{11.5cm}}Marianne should participate in the upcoming strike by stopping sales at her Ebsy shop for a week\\ Marianne should maintain a good relationship with Ebsy by keeping her shop open during the strike\\ Workers pool their savings to participate in the strike without losing income\\ Workers notify buyers of their situation to garner support\end{tabular}} &
  Worker \\ \cline{2-5} 
 &
  \multicolumn{3}{l|}{\begin{tabular}[c]{p{11.5cm}}Regulators impose a price ceiling on transaction fees allowed for locally crafted goods.\\ Regulators requires companies like Ebsy to implement transparent policies\end{tabular}} &
  Regulator \\ \hline
\end{tabular}
\end{table*}

\begin{table*}[]
\caption{Example instance of solution rankings}
\Description[A table of figures depicting four steps of an example instance of solutions-ranking.]{In the before ranking phase, solutions are sprawled out next to the board. After solutions are ranked on board, stickies appear in a vertical on the board, with most preferred solutions up ton. After participants added new solutions, green stickies appear on board. After post-ranking questions are answered, responses appear to the questions on the board (next to rankings).}
\centering
\begin{tabular}{|l|l|} 
\label{tab:rankings}
\includegraphics[width=8.5cm]{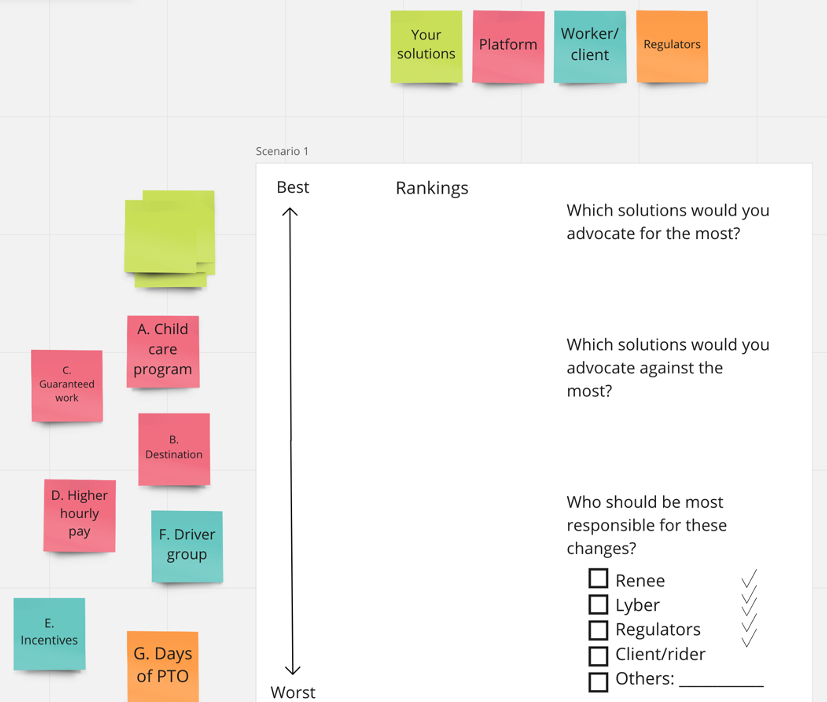} & \includegraphics[width=8.5cm]{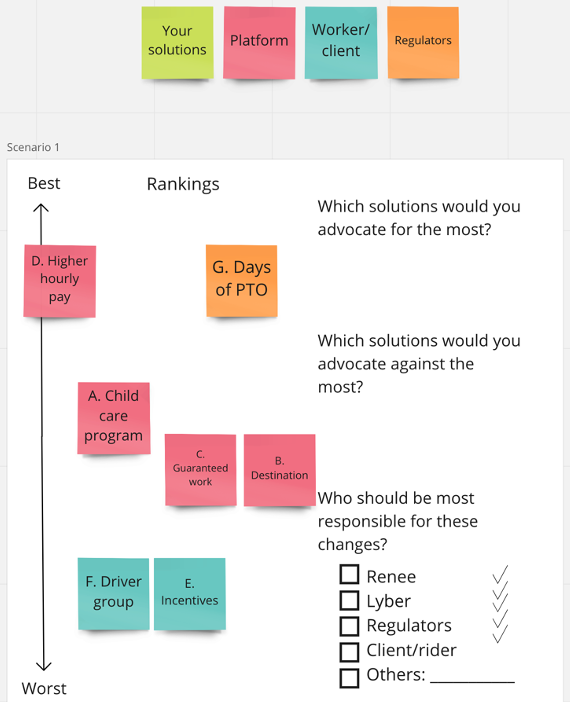} \\ 
\hline
Before ranking      & After ranking generated solutions  \\ 
\hline
\includegraphics[width=8.5cm]{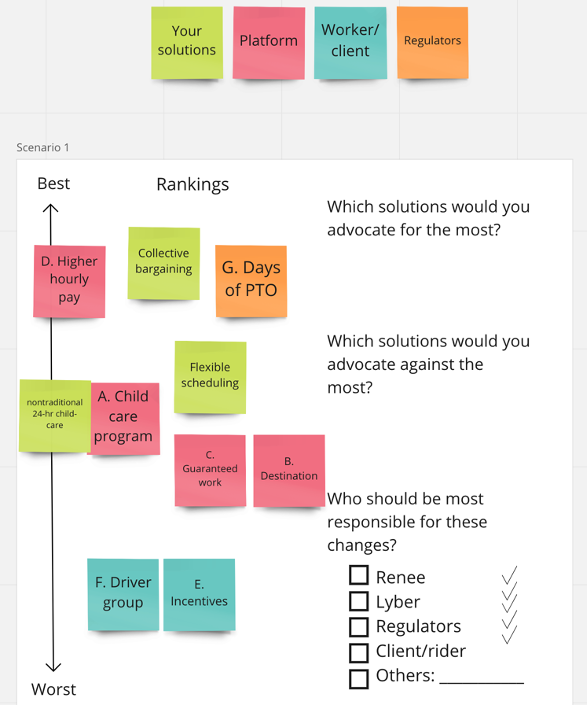} & \includegraphics[width=8.5cm]{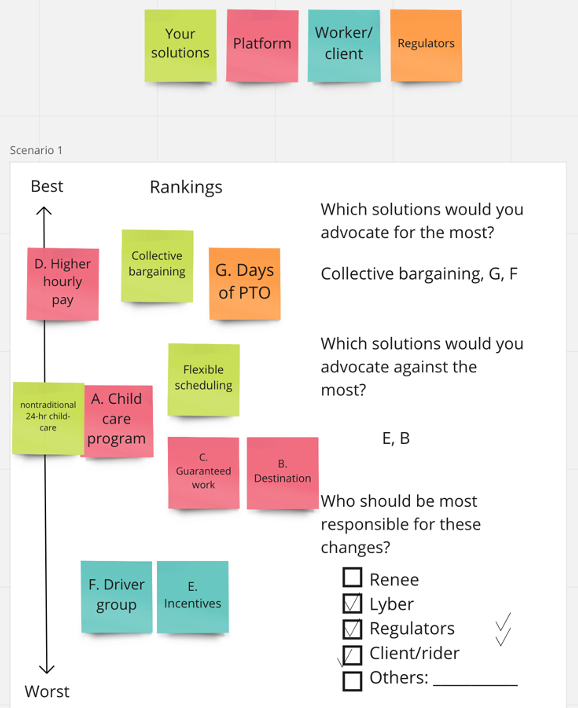} \\ 
\hline
Added new solutions & Answering post-ranking questions   \\
\hline
\end{tabular}

\end{table*}

\end{document}